\documentclass[default,iicol]{sn-jnl}% Default with double column layout

\usepackage{caption}
\usepackage{subcaption}
\usepackage[numbers]{natbib}
\usepackage{array, longtable, tabularx}% added long table
\usepackage{pdflscape}% added
\usepackage{hyperref}% had to be last in preamble
\usepackage{float}
\jyear{2022}%

\theoremstyle{thmstyleone}%
\theoremstyle{thmstyletwo}%

\theoremstyle{thmstylethree}%

\begin{document}

\title[DoubleU-NetPlus]{DoubleU-NetPlus: A Novel Attention and Context Guided Dual U-Net with Multi-Scale Residual Feature Fusion Network for Semantic Segmentation of Medical Images}


\author*[1]{\fnm{Md. Rayhan} \sur{Ahmed}}\email{rayhan@cse.uiu.ac.bd}

\author[2]{\fnm{Adnan Ferdous} \sur{Ashrafi}}\email{adnan@stamforduniversity.edu.bd}

\author[3]{\fnm{Raihan Uddin} \sur{Ahmed}}\email{raihanahmed95@stamforduniversity.edu.bd}

\author[1]{\fnm{Swakkhar} \sur{Shatabda}}\email{swakkhar@cse.uiu.ac.bd}

\author[1]{\fnm{A.K.M. Muzahidul} \sur{Islam}}\email{muzahid@cse.uiu.ac.bd}

\author[1]{\fnm{Salekul} \sur{Islam}}\email{salekul@cse.uiu.ac.bd}

\affil*[1]{\orgdiv{Department of Computer Science and Engineering}, \orgname{United International University}, \orgaddress{\street{United City, Madani Avenue}, \city{Badda}, \postcode{1212}, \state{Dhaka}, \country{Bangladesh}}}

\affil[2]{\orgdiv{Department of Computer Science and Engineering}, \orgname{Stamford University Bangladesh}, \orgaddress{\street{51, Siddeswari Road}, \city{Ramna}, \postcode{1217}, \state{Dhaka}, \country{Bangladesh}}}

\affil[3]{\orgdiv{Department of Electrical and Electronics Engineering}, \orgname{Stamford University Bangladesh}, \orgaddress{\street{51, Siddeswari Road}, \city{Ramna}, \postcode{1217}, \state{Dhaka}, \country{Bangladesh}}}


\abstract{Accurate segmentation of the region of interest in medical images can provide an essential pathway for devising effective treatment plans for life-threatening diseases. It is still challenging for U-Net, and its state-of-the-art variants, such as CE-Net and DoubleU-Net, to effectively model the higher-level output feature maps of the convolutional units of the network mostly due to the presence of various scales of the region of interest, intricacy of context environments, ambiguous boundaries, and multiformity of textures in medical images. In this paper, we exploit multi-contextual features and several attention strategies to increase networks' ability to model discriminative feature representation for more accurate medical image segmentation, and we present a novel dual U-Net-based architecture named DoubleU-NetPlus. The DoubleU-NetPlus incorporates several architectural modifications. In particular, we integrate EfficientNetB7 as the feature encoder module, a newly designed multi-kernel residual convolution module, and an adaptive feature re-calibrating attention-based atrous spatial pyramid pooling module to progressively and precisely accumulate discriminative multi-scale high-level contextual feature maps and emphasize the salient regions. In addition, we introduce a novel triple attention gate module and a hybrid triple attention module to encourage selective modeling of relevant medical image features. Moreover, to mitigate the gradient vanishing issue and incorporate high-resolution features with deeper spatial details, the standard convolution operation is replaced with the attention-guided residual convolution operations, which enables the network to achieve effective and relevant feature maps from a significantly increased network depth. Empirical results demonstrate that the proposed network accomplishes superior performance compared to other state-of-the-art methods on six benchmark datasets of diverse modalities. The proposed network achieves a dice similarity coefficient score of 85.17\%, 99.34\%, 94.30\%, 96.40\%, 95.76\%, and 97.10\% on DRIVE, LUNA, BUSI, CVCclinicDB, 2018 DSB, and ISBI 2012 datasets.
}




\maketitle

\section{Introduction}\label{sec1}

Medical imaging is a highly critical element in modern medical practice and biotechnology to undertake numerous diagnostic procedures, from wellness and screening to early diagnosis, clinical analysis, treatment selection, image-guided surgery, and subsequent follow-ups for continuous assessments of the patient's health condition~\cite{ liu2020survey}. It has become a crucial resource for physicians to understand and assess the disease. Moreover, it is essential to determine the efficacy of the treatment, allowing clinicians to better analyze a patient by creating a pictorial and functional representation of hidden physiological structures of body parts such as bones, organs, tissue, and blood vessels for clinical examination \cite{CNL-UNet, CMM-Net} and evaluate various cellular and molecular events. Non-invasive medical imaging techniques, such as X-ray, Computerized tomography (CT), Ultrasound, Colonoscopy, Dermoscopy, Microscopy,  Electrocardiogram (ECG), and Magnetic resonance imaging (MRI), can reveal crucial anatomical functionality-related information on diseases and anomalies within the body~\cite{CFPNet-M}.

Semantic medical image segmentation (MIS) is one of the significant areas of research in medical image analysis. In semantic segmentation, every distinct pixel in the image is assigned a distinct category, thus partitioning an image into a set of non-overlapping regions, which can also be regarded as a dense classification problem~\cite{MultiResUNet}. MIS refers to the process of distinguishing specific areas within a 2D or 3D medical image, which can facilitate the clinicians to study only the desired parts or region of interest (ROI) of the multi-modal medical images~\cite{MultiResUNet}. It is an essential preliminary step for any computer-aided diagnosis (CAD) system and often plays an integral role in both quantitative and qualitative analysis of medical images~\cite{CMM-Net}, such as segmentation of polyps~\cite{Msrf-net, Doubleu-net}, lung region~\cite{nishio2021lung, rahman2022improving}, brain tumors~\cite{Unet++}, retinal blood vessels~\cite{Bridge-Net}, cell nuclei~\cite{RIC-Unet}, cell contour~\cite{wang2021knowledge}, and breast ultrasound images~\cite{MCRNet}.

During the past decade, the vast majority of architectures created for semantic segmentation in various applications of computer vision (CV) and medical image analysis are based on deep neural networks (DNNs), such as fully convolutional networks (FCNs), coined by Long et al.~\cite{long2015fully} or encoder-decoder-based convolutional neural networks (CNNs) such as Seg-Net~\cite{Segnet}. The establishment of encoder-decoder-based CNNs achieved promising segmentation performance in CV and medical imaging. Nevertheless, U-Net, proposed by Ronneberger et al.~\cite{U-net}, made a significant breakthrough in the MIS task by incorporating the idea of skip connections between each symmetric layer of the encoder and decoder. Primarily, the encoder performs multiple convolutions and pooling operations to capture various representations of images, from low to high-level. It decreases the spatial dimensions of each layer and increases the number of channels. As the architecture goes deeper, more high-level feature maps, such as objects and various shapes, are captured. On the contrary, the decoder performs multiple up-sample and concatenation operations followed by convolution operations to predict the segmented mask. It increases the spatial dimensions while decreasing the channels.

In recent years, several variants of U-Net followed, such as U-Net++~\cite{Unet++}, MultiResU-Net~\cite{MultiResUNet}, LadderNet~\cite{LadderNet}, Attention U-Net~\cite{oktay2018attention}, R2U-Net~\cite{r2u-net}, CE-Net~\cite{Ce-net}, and KiU-Net~\cite{Kiu-net}. Even though these methods have improved the feature representation to some extraordinary level, they are still constrained by a number of significant drawbacks. Similar scale feature maps with various receptive fields that are generated from the convolution kernel have distinct semantic feature representations. The size of the receptive field in the convolutional kernel can affect network performance~\cite{MSU-net}. Most of the datasets have images where the ROI is of diverse shapes and sizes, for example, polyps in colonoscopy images. When the receptive field is too large, smaller targets can get disregarded, and on the other side, a smaller receptive field can capture redundant information. Hence, processing the image using convolution kernels with different receptive fields is vital for capturing the global contextual representation of features~\cite{Ce-net}. Because of the substantial loss of spatial information during encoding, it is usually challenging to reconstruct the details of low-level feature maps such as edges, dots, corners, and lines using orthodox de-convolution operations~\cite{Refinenet}. The resultant feature maps are sparse, resulting in a reduction in segmentation performance. Moreover, U-Net and its variants also suffer from semantic gaps in feature representations because of longer skip connections present between the corresponding encoder and decoder. Combining the two incompatible representations of feature sets of encoder and decoder blocks introduces inconsistency in the architecture's learning process. In order to reduce the semantic gaps and loss of spatial information during encoding and improve the high and low-level fusion of semantic information throughout the network, multiple U-Net-based architectures can be deployed to achieve state-of-the-art (SOTA) segmentation results~\cite{Doubleu-net}.

The attention mechanism concentrates solely on the most informative feature representations for a specific task without additional supervision, thereby penalizing less informative regions of the image and avoiding using similar feature maps in the network; thus, attention-based networks have recently been widely employed in MIS tasks. The channel-based attention mechanism is one of the most investigated attention mechanisms in literature. It exploits the inter-channel relations of features and focuses on desired object selection by actively re-calibrating each channel's  weight~\cite{guo2022attention}. Hu et al.~\cite{SEnet} initially presented the idea of channel attention and introduced SE-Net architecture. SE-Net utilizes the global average pooling mechanism to capture the global representations of contextual features. However, a simple global average pooling mechanism can fail to extract complex high-level intra-channel feature representations~\cite{gao2019global}. The spatial attention mechanism mainly focuses on relevant spatial regions of informative features. Nevertheless, integration of only SE-Net's attention has been found to be inadequate and sub-optimal in many MIS tasks~\cite{Cbam}. Woo et al.~\cite{Cbam} suggested the concept of a convolutional block attention module (CBAM), which is a sequential combination of these two attention mechanisms and can bring effective results for many CNN-based tasks. Oktay et al.~\cite{oktay2018attention} introduced a low-cost, lightweight attention gate mechanism to focus mainly on the selected ROIs while suppressing feature activations in non-ROIs. Recently various transfer learning techniques have been applied to the task of MIS due to their robustness and quick convergence mechanism~\cite{Doubleu-net, Ce-net, EANet}. It allows the pre-trained weights from one task to be utilized in different but related tasks.

In this paper, we extend and significantly improve the SOTA DoubleU-Net~\cite{Doubleu-net} architecture and propose a robust novel architecture that can effectively perform the MIS tasks of multi-modal domains by modeling global contextual information and high-level multi-scale semantic feature representations of pixels of varying receptive fields. EfficientNetB7~\cite{tan2019efficientnet} architecture is adopted through transfer learning as our backbone encoder module for extracting effective feature information. We incorporate a novel triple attention gate (TAG) mechanism in every skip connection to attend to selective inputs with high relevancy to the target region. To reduce the semantic gap issues of the skip connections of U-Net~\cite{U-net}, DoubleU-Net~\cite{Doubleu-net}, and other similar variants, we incorporate attention-guided residual (AG-Residual) convolution operations instead of regular convolutions. We also design a multi-kernel residual convolution (MKRC) module to acquire high-level global contextual features. The MKRC block extracts fine-grained contextual information of higher levels from images with various levels of receptive fields such as $1 \times 1$, $3 \times 3$, $5 \times 5$, and $7 \times 7$. The receptive field of a CNN usually refers to the size of the kernel. The generated feature maps from the MKRC block are then passed through the newly designed squeeze and excitation-based atrous spatial pyramid pooling (SE-ASPP) module~\cite{Deeplab} to extract high-resolution relevant feature maps for effective learning of the proposed model. In addition, inspired by the CBAM architecture~\cite{Cbam}, we also integrate a hybrid triple attention module (TAM), which performs features refinement through parallel execution of spatial attention, modified channel-based attention mechanism, and squeeze and excitation-based attention to capturing relevant spatial regions of the higher-level global contextual features and inter-dependencies among different channels, respectively. 

Overall, the main contributions of this work can be summarized as follows:
\begin{itemize}
  \item A robust EfficientNetB7 encoder backbone-based segmentation framework, referred to as DoubleU-NetPlus, is proposed to enhance the semantic segmentation performance for medical images.
  \item A newly proposed multi-kernel residual convolution module, which expands the field of view representation of heterogeneous, semantic global contextual features at different scales.
  \item A modified hybrid triple attention module, which performs an aggregation of spatial and channel-based attention and squeeze and excitation-based attention, thus, improves the channel inter-dependencies and inter-spatial relationships of the high-level feature maps.
  \item We integrate a novel lightweight triple attention gate module at the decoder side of each network to highlight salient features from the skip connections.
  \item Embedding of features re-calibration through squeeze and excitation operation in the attention-based atrous spatial pyramid pooling mechanism.
  \item We demonstrate the effectiveness of the proposed method on six publicly available benchmark datasets, and comparative analysis show that DoubelU-NetPlus outperformed the SOTA medical segmentation models.
\end{itemize}

\section{Related Works}
This section provides a brief summary of the research pertaining to MIS techniques, including context-aware segmentation, attention-guided segmentation, and stacked multi-U-Net techniques.

\subsection{Context-aware segmentation}
Contextual information from multiple levels of a network plays a significant role in the performance of any CNN-based MIS model. Xie et al.~\cite{Xie2022} proposed a context hierarchical integrated network (CHI-Net), which introduced a dense dilated convolution module for gathering features from four cascaded branches of hybrid dilated convolutions. The authors also introduced a stacked residual pooling module that uses multiple effective fields of view. Residual dilated convolution was utilized in the encoder part of the network to capture multi-level hierarchical features. Gu et al.~\cite{Ce-net} used a context encoder network (CE-Net) that utilizes a pre-trained ResNet-34 as the encoder module. The authors integrate a context extractor module consisting of a dense atrous convolution block and a residual multi-kernel pooling block. Al masni et al.~\cite{CMM-Net} applied a Contextual Multi-Scale Multi-Level Network (CMM-Net) by fusing the global contextual features of different spatial scales in the encoding part of the U-Net. The authors also used a dilated convolution module that expanded the receptive field with different rates depending on feature maps network sizes.

Xiao et al.~\cite{Xiao2022} introduced a deep residual contextual and sub-pixel convolution network (RC-SPCNet) for the segmentation of neuronal structure. The encoding section of the U-Net included residual-convolution blocks along with summation-based skip connections, and the decoding section was deployed with sub-pixel convolutional layers. Lifted multi-cut was used for optimizing the output for reconstruction results. Lou et al.~\cite{MCRNet} introduced an inverted residual pyramid block and a context-aware fusion block in a new U-net architecture. The authors deployed a multi-level context refinement network (MCRNet) using these two context refinement blocks into a U-net architecture in a multi-level manner. Finally, Wu et al.~\cite{WU2021102025} proposed a new U-Net architecture comprising three new modules, namely, a scale-aware feature aggregation module, an adaptive feature fusion module, and a multi-level semantic supervision module.

Recently various transformer-based architectures have been effectively used in the MIS task too. By modeling global context-based features effectively, architectures like Swin-UNet~\cite{Swin-unet}, Ds-TransUNet~\cite{Ds-transunet}, and UNETR~\cite{Unetr} achieved SOTA results on MIS tasks of diverse modalities.

In all of the above works, the authors tried to extract multi-scale representations so that gaps in semantics between the encoder and decoder features could be reduced. Although, these readjustments in many a case introduced over-fitting problems \cite{MDU-Net}, which resulted in not so much significant rise in evaluation metrics.

\subsection{Attention-guided segmentation}
Over the years, with the successful application of many computer vision-oriented tasks, various attention mechanisms have been increasingly applied to the field of MIS. Wang et al.~\cite{EANet} proposed an iterative edge attention network (EANet) where the authors integrated the edge-attention preservation (EAP) module along with a dynamic scale-aware context (DSC) module. The authors employ the VGG-19~\cite{vgg} feature encoder. The EAP module captures edge-related attention information such as background noise and shape by preserving the low-level local edge features. The gated convolutional blocks (GCB) interleaved with some residual blocks in the EAP module allow the edge stream to solely analyze boundary-related data.

Zhao et al.~\cite{SCAU-net} proposed a MIS architecture where the authors apply spatial and squeeze and excitation networks (SE-Net) to focus mainly on the initial low-level feature maps and channel inter-dependencies in the high-level feature maps in the bottleneck part of the network. Wang et al.~\cite{SAR-U-Net} incorporate the SE attention mechanism in the encoder part of the network to adaptively extract the feature maps and the ASPP module to capture the context-based semantic information from the extracted feature maps at multiple scales. SE-Net is also incorporated by Li et al.~\cite{Res2Unet}, where the authors use Res2Net~\cite{Res2net} as the encoder backbone. The extracted features are grouped by channels, and convolution operations are performed on each group separately. SE-Net is integrated to learn the relationship between groups and re-calibrate the channel weights to focus on the target object.

Gao et al.~\cite{gao2021multiscale} proposed a multi-scale fused network that employs two attention mechanisms, additive channel attention and additive spatial attention in the skip connections, which utilize high-level features to prune the responses of low-level features in both channel and spatial dimensions. It improves the learning of the superior spatial relationship between adjacent pixels and inter-dependencies between channels. Yeung et al.~\cite{Focus} proposed an attention-gated U-Net architecture that employs a new attention module named Focus Gate and combines spatial and channel-based attention with a focal parameter to regulate the degree of background suppression. The focus gate utilizes the gating signal to refine incoming signals from the encoding network as long-range skip connections, indicating selected image features and regions included in the decoding network.

Tomar et al.~\cite{Fanet} introduced a new attention-based mechanism named FANet, which combines the feature maps from the current training epoch with the prior epoch mask. The prior epoch mask provides hard attention to the learned feature maps at different convolutional layers. Han et al.~\cite{ConvUNeXt}, in their proposed ConvUNeXt architecture, utilized the ConvNeXt~\cite{ConvNeXt} as the encoder backbone along with the attention gate mechanism in every skip connection. Tong et al.~\cite{ASCU-Net} also utilized the lightweight attention gate mechanism in the decoder part of the network. The feature map generated by the attention gate module is processed by the channel and spatial attention modules in parallel, whose outputs are combined to produce the final feature maps.

Though all the aforementioned attention-based methods achieved reasonable performance in the MIS tasks, they still face difficulties in achieving SOTA segmentation performance in terms of diverse shapes and subjects, especially in the ultrasound image and retinal modalities.

\subsection{Stacking/Cascading of multiple U-Nets}

Another popular method to improve the MIS performance is to stack multiple U-Net architectures together in a \textit{k}-cascading U-Net format, where \textit{k} refers to the number of sub-U-Nets~\cite{zhang20222k}. For example, DoubleU-Net~\cite{Doubleu-net}, with two U-Net architectures stacked on top of each other. Ghosh et al.~\cite{ghosh2018stacked} proposed the idea of incorporating dilated stacked U-Nets for semantic scene segmentation. In another work, Ding et al.~\cite{ding2019stacked} utilize a series of U-Nets stacked together for brain tumor segmentation. In addition, a two-level nested U-net structure with encoders and decoders comprised of U-Net structured modules has been constructed~\cite{qin2020u2}. Furthermore, W-shaped networks have been established in recent years. W-Net~\cite{W-net} functions by concatenating two U-Nets into an autoencoder format, one for encoding and one for decoding, and achieves satisfactory results in unsupervised image segmentation tasks.

All of the above-mentioned architectures connect two or more U-Nets together and can therefore extract a separate group of features using the same set of original features. However, the challenge is that the same features may be extracted repeatedly, which can degrade the network's efficiency~\cite{zhang20222k}.

\section{Proposed Method}

\begin{figure*}[!htb]
    \centering
    \includegraphics[height = 17 cm, width=0.80\textwidth]{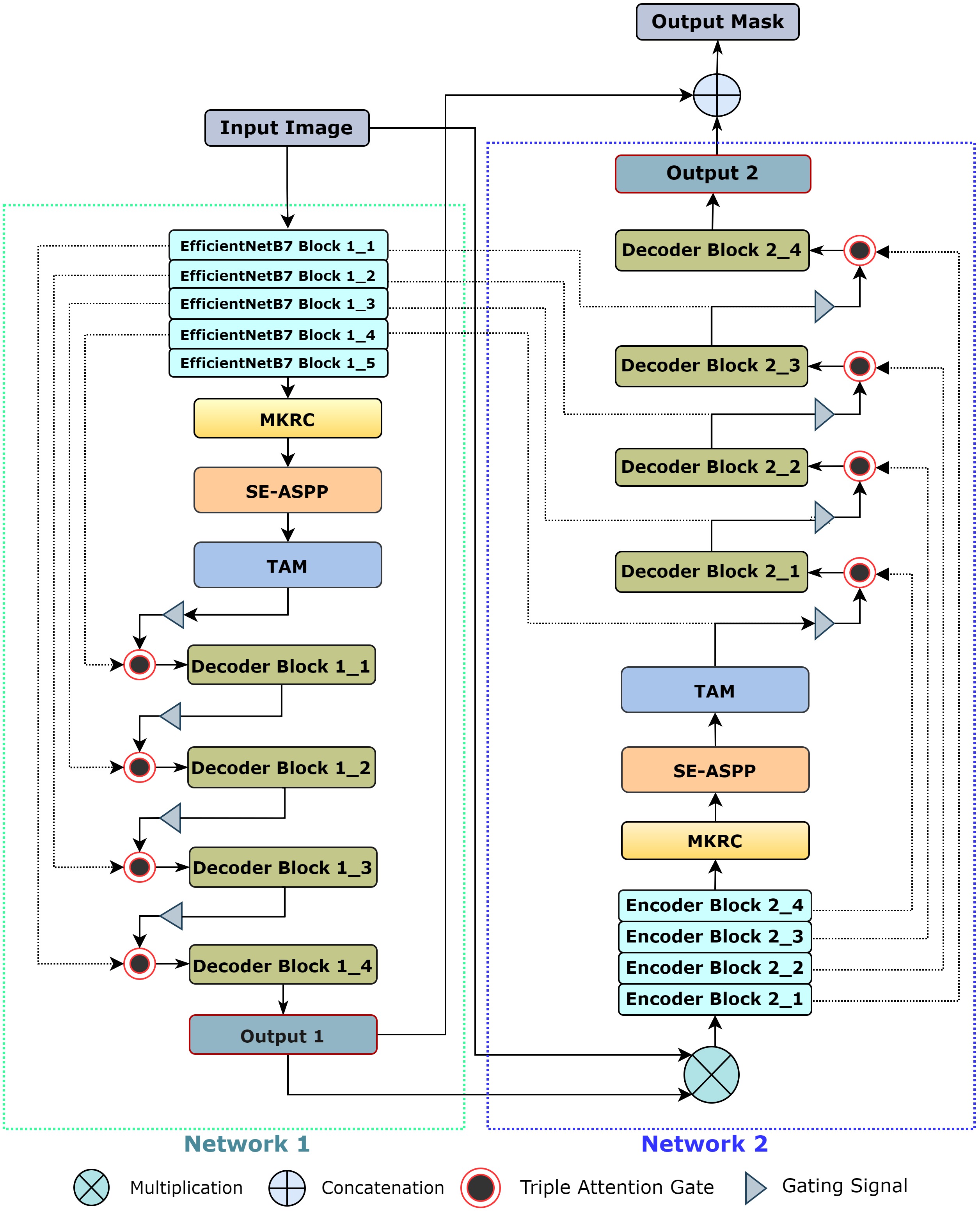}
    \caption{{\scriptsize Composition of proposed DoubleU-NetPlus architecture.}}
    \label{fig:dual_u_net}
\end{figure*}

In this section, we describe the architecture of the proposed segmentation network and the details of the constituent modules. Firstly, the architecture of the DoubleU-Net~\cite{Doubleu-net} network is briefly described, and then we elaborately describe the proposed architecture and the incorporated modules in it. The proposed architecture is demonstrated in Fig.~\ref{fig:dual_u_net}. 

\subsection{Overview of DoubleU-Net architecture}
DoubleU-Net~\cite{Doubleu-net} is an encoder-decoder architecture comprising two U-Net networks stacked on top of each other. There are two encoders and two decoders in the DoubleU-Net architecture. In the first U-Net architecture, VGG-19~\cite{vgg} is incorporated as the backbone of the first encoder, which is pre-trained on ImageNet~\cite{Imagenet}. The decoder of the first U-Net architecture is built by performing the upsampling of the feature maps, then concatenating with the corresponding skip connections, and lastly, two regular convolution operations of $3 \times 3$ followed by batch normalization, ReLu, and squeeze and excitation operation. In order to utilize more high-level semantic information efficiently, the authors placed the second U-Net at the bottom of the first U-Net. The encoder of the second U-Net is formed by performing consecutive convolution and max-pooling operations. The decoder of the second U-Net is similar to the decoder of the first U-Net. The results generated by the DoubleU-Net architecture outperformed several MIS algorithms by a significant margin in four benchmark datasets. Despite achieving significant performances, DoubleU-Net lacks effectiveness in the skip connections of the network~\cite{sang2021ag}, limiting the precise flow of information throughout the network. Moreover, it does not fully exploit the high-level feature maps from varying receptive fields, which can increase the results further. A further shortcoming of DoubleU-Net is its outdated VGG-19 encoder backbone, which can be replaced by a more recently proposed deeper architecture like EfficientNetB7~\cite{ConvUNeXt}. Hence, we select DoubleU-Net as our basic architecture for further enhancement.

\subsection{Overview of the proposed DoubleU-NetPlus architecture}

We performed enhancements in both the networks of the DoubleU-Net architecture by deploying the EfficientNetB7 architecture as the encoder one backbone for extracting multi-scale information. In all the skip connections, we employ a novel triple attention gate (TAG) module to selectively attend to the significantly relevant features in the decoder while suppressing irrelevant feature representations. In comparison to high-level feature information, low-level feature information tends to contribute less to network performance and use a lot of computational resources, also pointed out by~\cite{song2022attention,wu2019cascaded}. As demonstrated in Fig. \ref{fig:dual_u_net}, to capture more effective multi-scale high-level contextual encoder information and pass it to be decoded by the decoder in the bottleneck/bridge of each encoder-decoder network, we design and embed the multi-kernel residual convolution (MKRC) module, modified squeeze and excitation-based atrous spatial pyramid pooling (SE-ASPP) module, and triple attention module (TAM) sequentially. Deeper networks considerably enhance the performance of the model. However, an increase in the depth of the network might occur in vanishing or exploding gradient problems~\cite{he2016deep, SAR-U-Net}. In order to address this issue and reduce the semantic gaps between the feature representations of the encoder and decoder, we utilize shortcut connections between layers in the residual learning paradigm. We have performed attention-guided residual (AG-Residual) convolution operations (see Fig.~\ref{fig:SSE-Residual}) in the encoder of the second network and decoders of both networks. The motivation behind deploying two multi-contextual attention-guided residual U-Net architectures is that the output feature maps of network one are not fully explored~\cite{zhang20222k}. We can enhance it by capturing the unexplored high-level multi-contextual information from the generated output feature maps of network one by multiplying it with the original input image and processing them together again in the second network to capture more semantic information. 

\subsection{Encoder and Decoder}

\begin{figure*}[!htb]
    \centering
    \includegraphics[width=0.90\textwidth]{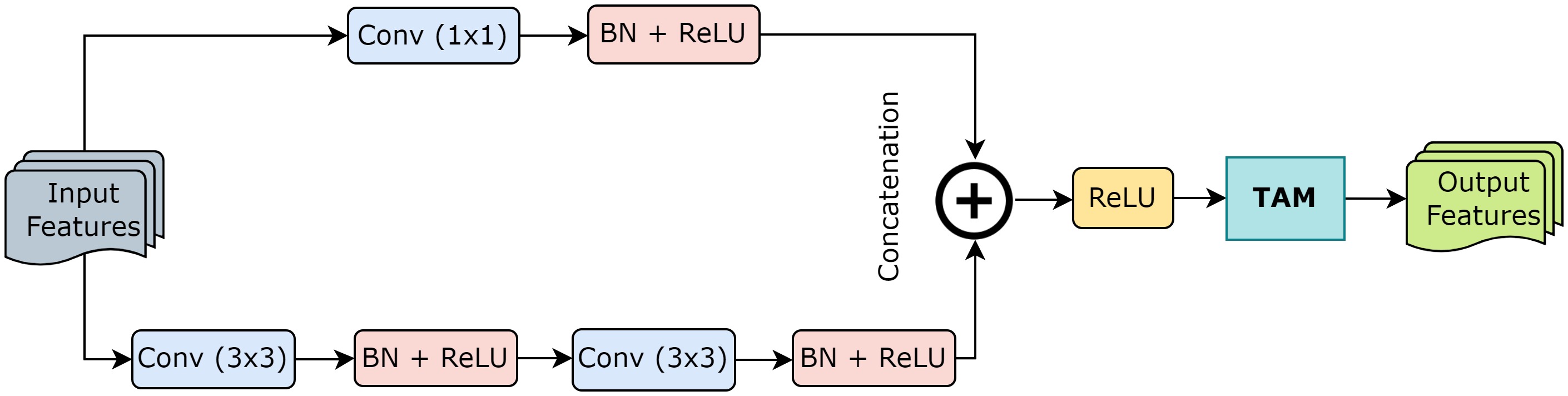}
    \caption{{\scriptsize Composition of the AG-residual convolution module.}}
    \label{fig:SSE-Residual}
\end{figure*}

The encoder portion of a U-Net is responsible for condensing the spatial information by each level. While it does so, the number of inputs halves, and the number of channels doubles. Consequently, we are left with highly condensed feature information that needs to be passed on to be decoded by the following levels. In our proposed DoubleU-NetPlus architecture, we utilize the EfficientNetB7 pre-trained architecture as the backbone for the encoder of network one using the transfer learning method, whereas the encoder in network two is built by performing two residual convolutions of  $3 \times 3$ followed by spatial and channel attentions. In the first encoder, we chose the EfficientNetB7 architecture mainly because of its higher accuracy and increased network depth. The deployment of EfficientNetB7 as the encoder of the first network gives the network effective feature extraction capability that the decoder of the first network can employ to generate extremely precise segmentation maps~\cite{siddique2021recurrent}. EfficientNetB7 implements a mobile inverted bottleneck convolution with an injected SE-Net~\cite{SEnet} block, which can attend to relevant features. By utilizing shortcuts directly between bottlenecks, which connect a significantly less number of channels than expansion layers, and depth-wise separable convolution, which effectively reduces computing cost compared to traditional layers. It performs more effectively by uniformly scaling the network's resolution, depth, and width, resulting in improved performance. Hence, deploying an EfficientNetB7 encoder enables us to have a contracting path that is significantly deeper and can perform effective contextual feature extraction of medical images. Each encoder block of the second encoder executes AG-residual convolution operations, as illustrated in Fig. \ref{fig:SSE-Residual}. The AG-residual convolution module performs two $3 \times 3$ convolution operations, each of which is followed by batch normalization and ReLU. The batch normalization decreases the internal covariant shift and regularizes the model~\cite{Doubleu-net}, while ReLU introduces non-linearity to the architecture. A shortcut residual connection is added with a $1\times1$ convolution of the input features to provide an identity mapping of features, followed by batch normalization and ReLU operations. Features from the $3 \times 3$ convolution operations and $1 \times 1$ shortcut connection are concatenated, followed by another ReLU operation. The generated feature maps are then passed to the TAM module, which performs both spatial and channel-based attention as well as  squeeze and excitation-based attention on the features to focus more on the relevant feature maps. Then we perform a max-pooling operation with a $2 \times 2$ window and stride of $2 \times 2$ to reduce the spatial dimension of the feature maps.

As shown in Fig. \ref{fig:dual_u_net}, there are two decoders in the architecture, one in each network. Each input feature is passed to the gating signal module, which captures high-level feature representations from the immediate lower part of the network. Then, each block in the decoder applies a $2\times2$ up-sampling of bi-liner interpolation to each input feature, hence doubling the dimension of the input feature maps. The generated feature maps are then passed to the attention gate module, which takes the skip connections and the gating signal as inputs and perform additive soft attention on these two feature maps, and the network learns to attend to the desired ROI while suppressing feature activation in irrelevant areas as the training proceeds. Then we concatenate the up-sampled feature maps with the output feature maps of the attention gates. The concatenated feature maps are then passed to the AG-residual module for attention-based convolution operations. Every skip connection in the proposed model passes through the attention gate. In the first decoder, we only employ attention-gated-skip connections from the first encoder of network one. However, in the second decoder, we use attention-gated-skip connections from both the encoders from networks one and two. This procedure maintains spatial resolution and improves the output feature maps' quality without focusing on irrelevant regions. Similar to the DoubleU-Net architecture~\cite{Doubleu-net}, the final step is applying a convolution layer with a sigmoid activation function to construct the mask for the modified U-Net.

\subsection{Multi-kernel residual convolution module}

\begin{figure*}[!htb]
    \centering
    \includegraphics[width=0.90\textwidth]{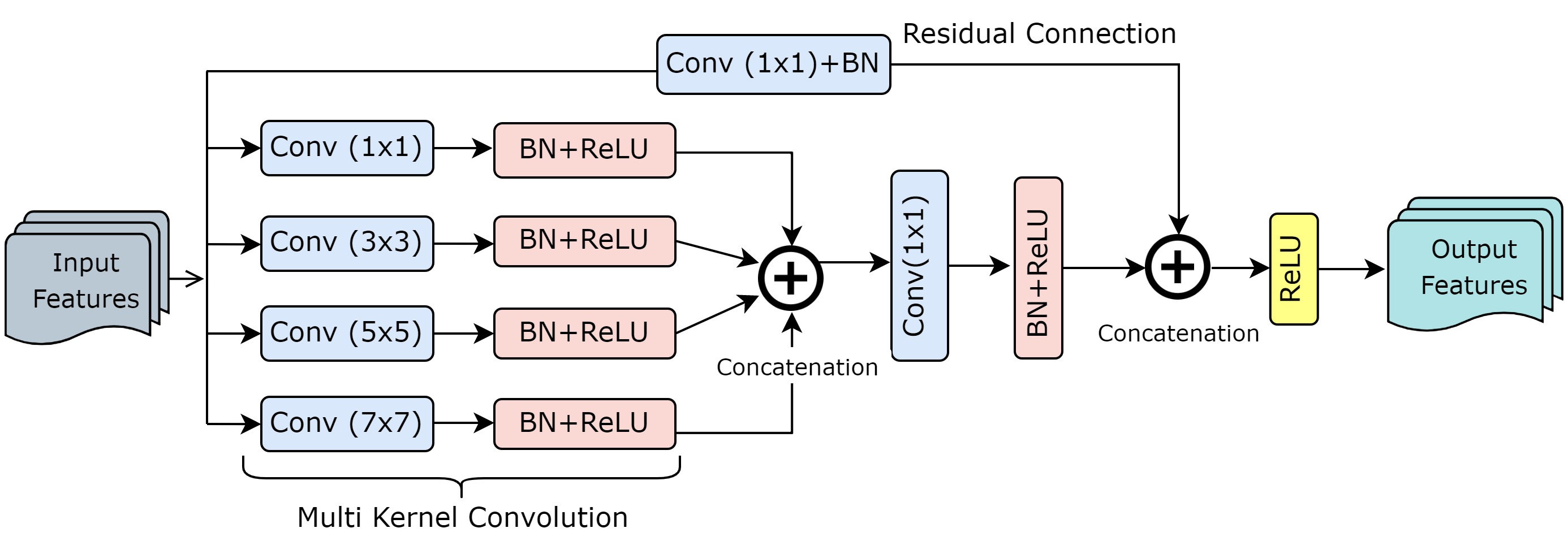}
    \caption{{\scriptsize Composition of the Multi-kernel residual convolution (MKRC) module.}}
    \label{fig:MKRC}
\end{figure*}

One of the challenges in MIS is the larger variation in the size and shape of an object in the medical image. Hence, to achieve effective results in the MIS task, it is necessary to extract high-level multi-scale contextual features through different receptive fields. In our proposed architecture, we applied an inception-inspired~\cite{inception} multi-kernel residual convolution (MKRC) module in both of the bottlenecks of networks one and two, which helps reduce saturation and degradation in the learning gradient. The proposed MKRC module is demonstrated in Fig. \ref{fig:MKRC}. The MKRC module expands the field of view representation of heterogeneous features for more effective and robust learning of the model. The module consists of multiple parallel convolution layers with different kernel sizes of ($1 \times1 $), ($3 \times3 $), ($5 \times5 $), and ($7 \times7 $), respectively. Increasing the kernel size in the convolution layers enables the networks to extract a more robust feature representation from multi-scale receptive fields, causing them to modulate the learning of features differently for each block. The next step after each convolution layer is a batch normalization layer and a ReLU activation function. After that, all four feature maps are concatenated together, which leaves us with information on every relevant receptive field. Next, we feed the concatenated feature maps to a ($1 \times 1$) convolution followed by batch normalization and ReLU. Next, we integrate a residual shortcut connection, also known as identity mapping, passed through a ($1\times 1$) convolution and batch normalization and perform concatenation with the previously generated feature maps. A ReLU activation is performed next. The resulting feature maps are then processed through a modified SE-ASPP module that expands the field-of-view representation of features to encompass a broader context.

\subsection{Squeeze and excitation-based atrous spatial pyramid pooling module}
\begin{figure}[!htb]
    \centering
    \includegraphics[width=0.45\textwidth]{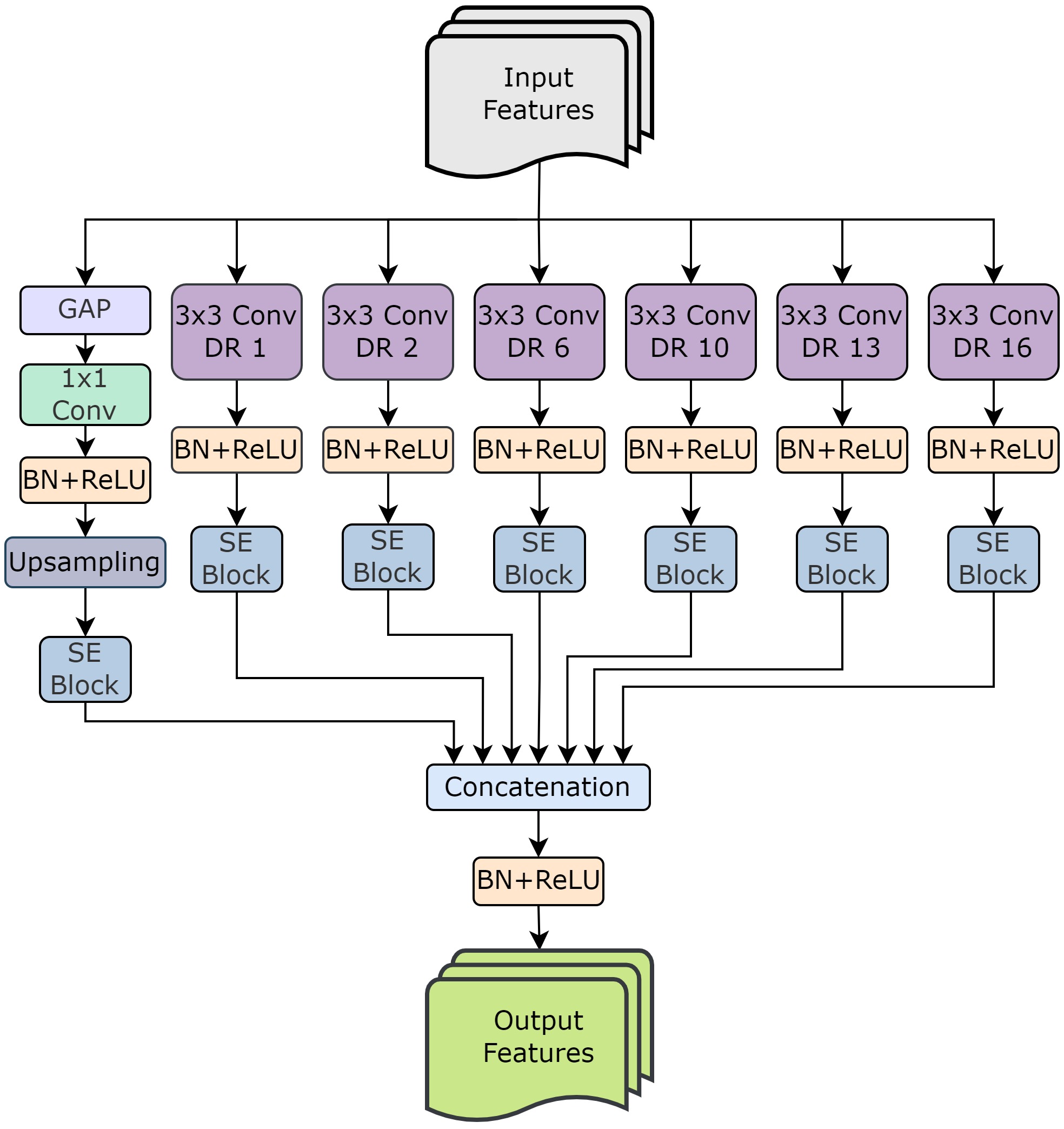}
    \caption{{\scriptsize Composition of the squeeze and excitation-based atrous spatial pyramid pooling (SE-ASPP) module.}}
    \label{fig:ASPP_module}
\end{figure}

Atrous spatial pyramid pooling (ASPP) introduced by Chen et al.~\cite{Deeplab} allows us to effectively enlarge the filters' field of view to include multi-scale contextual representation of semantic features by parallel atrous convolution layers with different dilation rates. It can efficiently mitigate the issue of reduced spatial resolution resulting from repeated down-sampling in the encoder~\cite{SAR-U-Net}. We modify the ASPP module and propose a new SE-ASPP module by embedding the squeeze and excitation networks (SE-Net) to the increased and enlarged field of view of the convolution filters. The structure of the SE-ASPP module is demonstrated in Fig. \ref{fig:ASPP_module}. We have utilized a deeper set of dilated convolutions in the SE-ASPP module in order to capture more robust and expanded representations of features from the MKRC module. The dilation rates utilized in the seven parallel convolution layers of the SE-ASPP module are 1, 1, 2, 6, 10, 13, and 16, respectively. We apply the squeeze and excitation network to effectively re-calibrate and refine the acquired features through different dilation rates. All the feature maps from the SE-Net modules of each branch of the SE-ASPP network are concatenated together, and a ($1 \times 1$) convolution operation is performed on the concatenated feature maps, followed by batch normalization and a ReLU activation function. The SE-ASPP module captures efficient and relevant semantic information at multi scales. The generated feature maps are then passed to the hybrid triple attention module (TAM) for further processing.

\subsection{Hybrid Triple attention module}

\begin{figure}[!htb]
    \centering
    \includegraphics[width=0.45\textwidth]{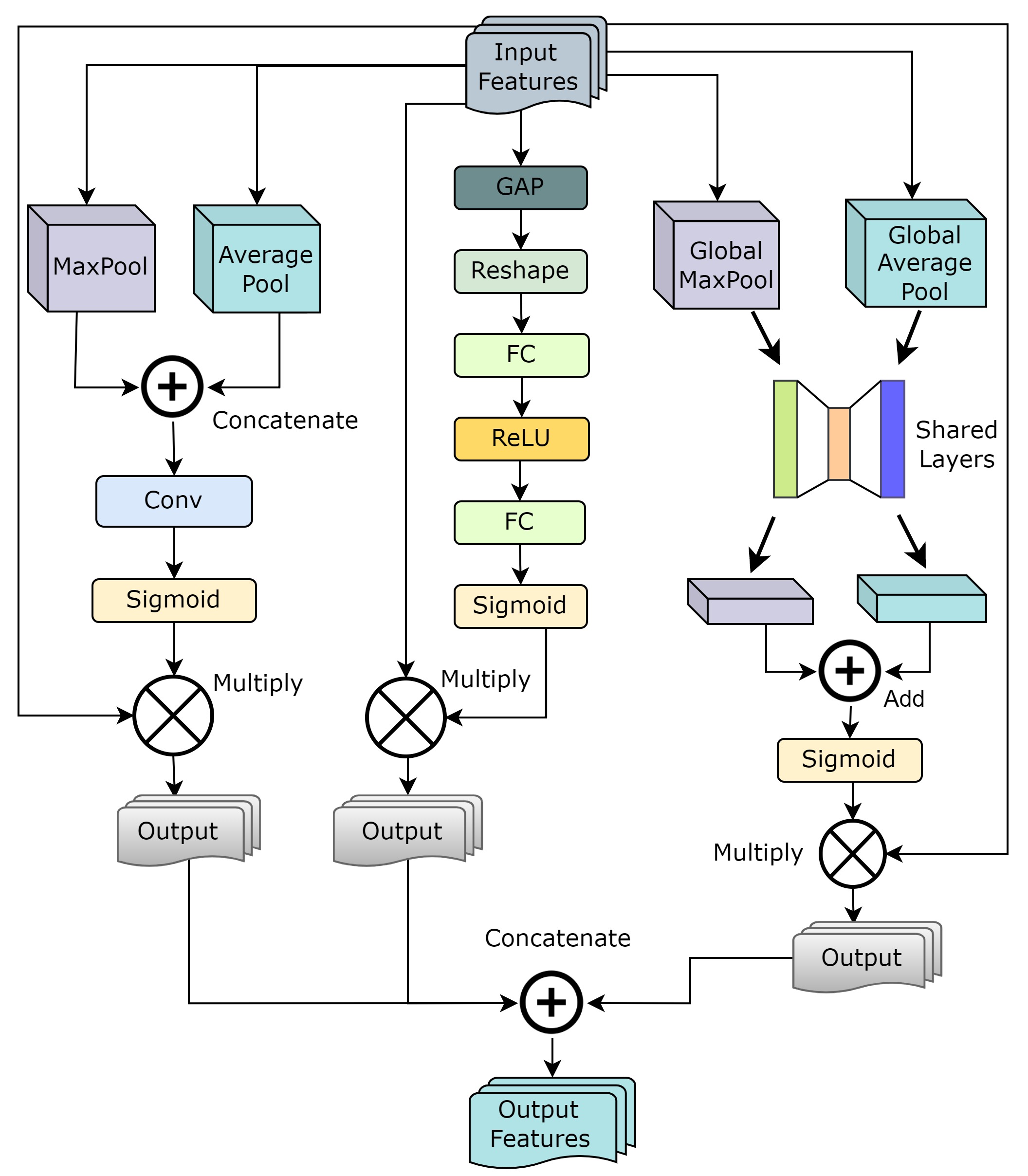}
    \caption{{\scriptsize Composition of the triple attention module (TAM).}}
    \label{fig:SCA_module}
\end{figure}

The hybrid triple attention module (TAM) performs effective attention-based feature refinement and extends the concepts introduced by CBAM~\cite{Cbam} and Focus U-Net~\cite{Focus}. As shown in Fig. \ref{fig:SCA_module}, it performs a feature fusion through parallel processing of squeeze and excitation networks~\cite{SEnet}, modified channel-based attention, and spatial attention mechanisms. We utilize these attention mechanisms to fully explore the high-level inter-spatial relationship of relevant features and effective inter-channel relationships. By adjusting the weight of each channel, SE-Net offers channel-based attention that can improve the channel inter-dependencies and can be seen as an object selection process while suppressing noise. However, SE-Net performs only global average pooling operations to perform channel-based attention. Later CBAM~\cite{Cbam}  suggested that these features could be sub-optimal and suggested using max pooling operations for modeling improved channel inter-dependencies. As illustrated in Fig. \ref{fig:SCA_module}, to achieve effective channel-based attention, we extend the ideas of CBAM and employ initial global average pooling and global max pooling operations along the channel axis, followed by concatenation and sigmoid activation to generate efficient feature descriptor that helps to determine where to highlight or suppress along the channel axis.
Through the spatial attention mechanism, the architecture focuses on the location of high-level feature maps of the target regions. In conjunction with channel-based attention, spatial attention module aggregate features along the channel axis~\cite{SEnet,Cbam,Focus}. We utilized the CBAM implementation of spatial attention by establishing two distinct channel contexts using average and max pooling along the channel axis, following spatial re-calibration using a kernel of size 7. Similar to modified channel-based attention, we experimented by incorporating initial global average pooling and global max pooling operations in the spatial attention module; however, the performance did not improve, and hence opted to use the original implementation of CBAM.

\subsection{Triple attention gate module}

\begin{figure*}[!htb]
    \centering
    \includegraphics[width=0.85\textwidth]{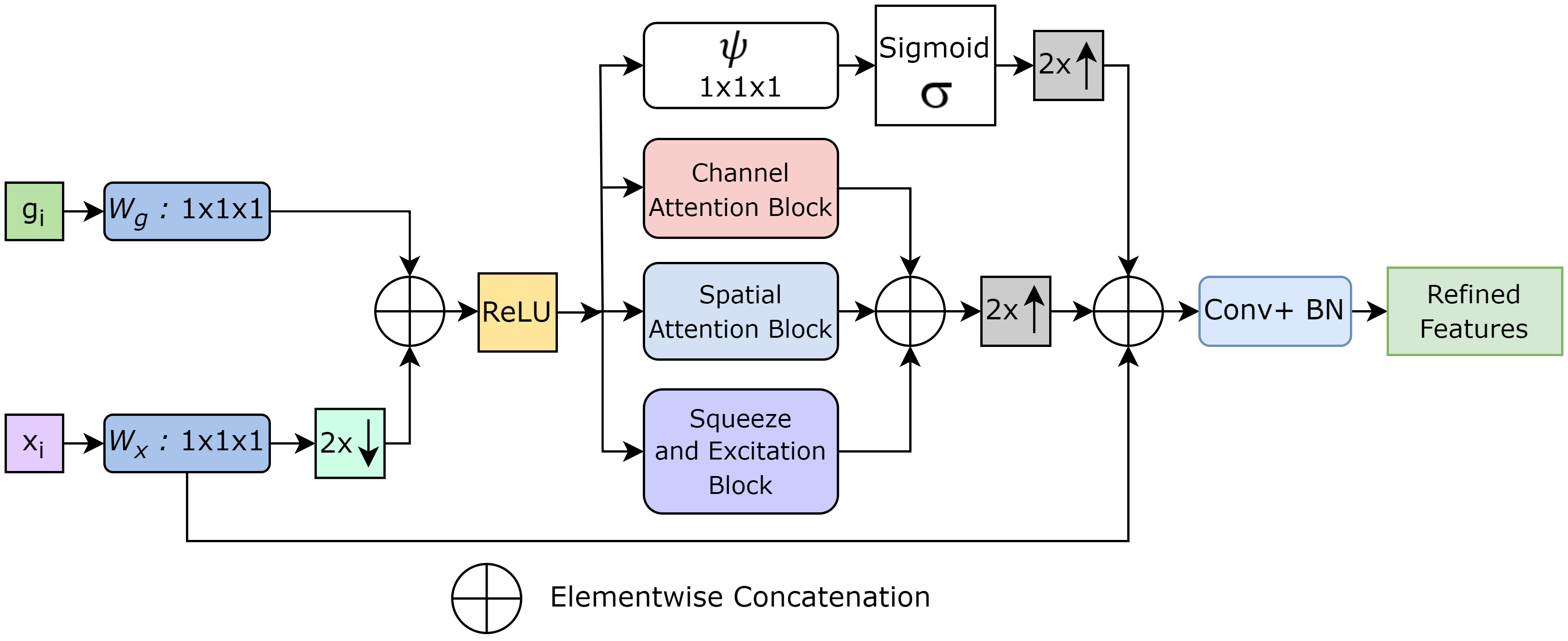}
    \caption{{\scriptsize Composition of the Triple Attention Gate (TAG) module.}}
    \label{fig:Attention_gate}
\end{figure*}

 Having introduced the SE-Net, channel-based attention, and spatial attention modules in the previous subsection, we describe the structure of the triple attention gate (TAG) module. Due to the lightweight design of the attention gate module, it significantly improves the model's representation ability without significantly increasing the computing cost or the number of model parameters~\cite{oktay2018attention}. Here, similar to the attention gate and focus gate~\cite{Focus} modules, we introduce a novel triple attention-gated deep neural network named the TAG module, which performs parallel implementation of channel attention, spatial attention, and squeeze and excitation-based attention mechanisms into a single TAG module to encourage selective learning of efficient, relevant features. The TAG module takes two inputs, as shown in Fig. \ref{fig:Attention_gate}, one is the gating signal from the one-step lower levels, which has a better representation of features such as edges, texture, and dots through training, and the other is the corresponding skip connection at that level, having a better representation of the spatial information. First, the gating signal and skip connection are resized to matching dimensions, and then they are combined through element-wise addition followed by non-linear activation (ReLu) and create attention coefficients. After that, the attention coefficients are passed through the channel, spatial, squeeze, and excitation-based attention modules and are then concatenated together to produce effective refinement of the relevant features. Next, we perform a $2 \times 2$ up-sampling operation to match the dimensions from the output of the $1 \times 1$ convolutions, followed by sigmoid operations, and $2 \times 2$ up-sampling performed on the output of the previously mentioned non-linear activation function. The aligned weights get larger, and the unaligned weights get relatively smaller. The spatial contextual information of ROIs is captured by concatenating the original skip connection by the generated attention coefficients. Hence, the vector gets scaled based on its relevance.

\section{Experimental analysis}
\label{sec:sample2}

\subsection{Datasets}
\begin{figure*}[!htb]
     \centering
     \begin{subfigure}[b]{0.16\textwidth}
         \centering
         \includegraphics[width=\textwidth]{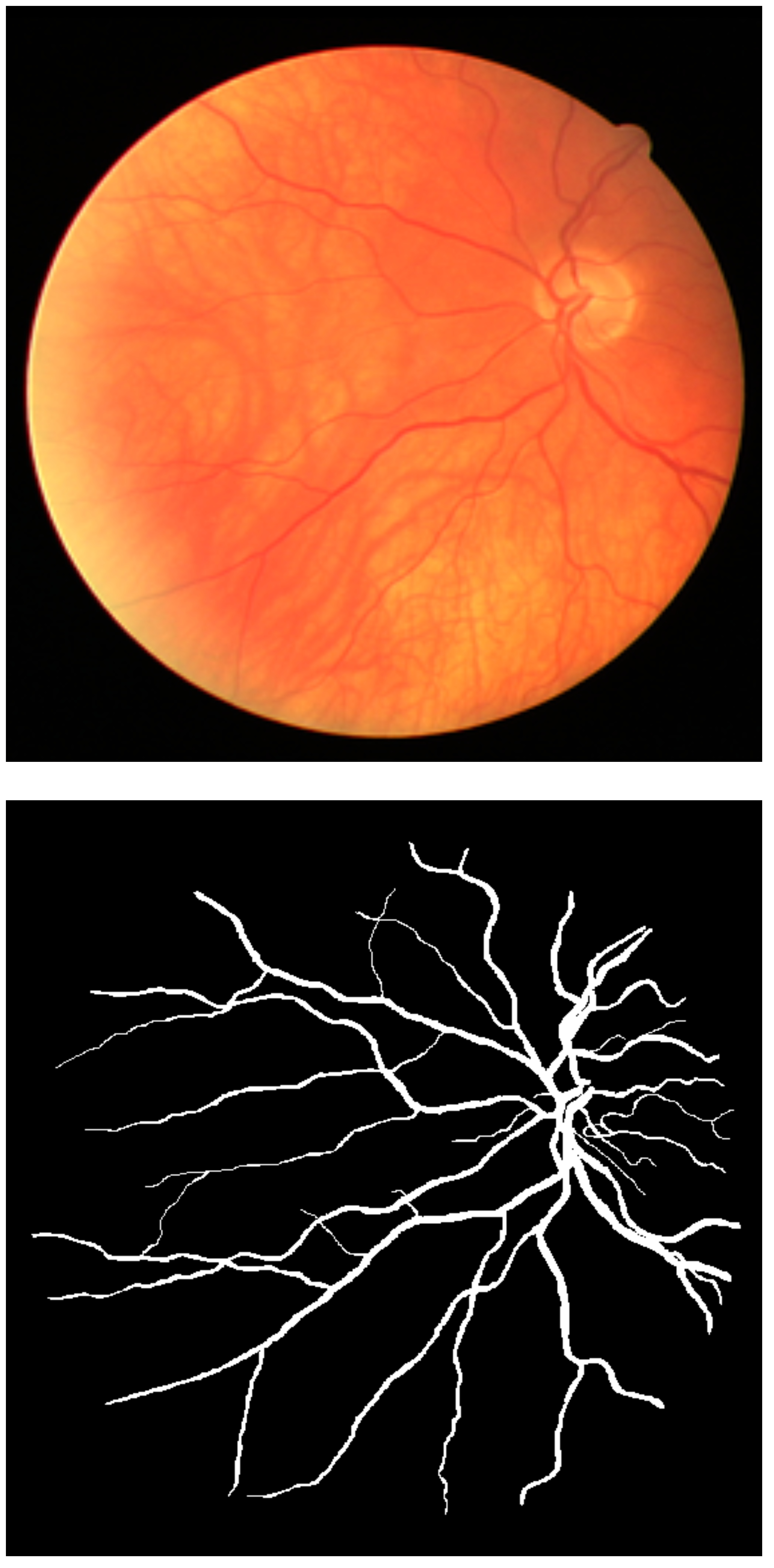}
         \caption{DRIVE}
         \label{fig:drive}
     \end{subfigure}
     \hfill
     \begin{subfigure}[b]{0.16\textwidth}
         \centering
         \includegraphics[width=\textwidth]{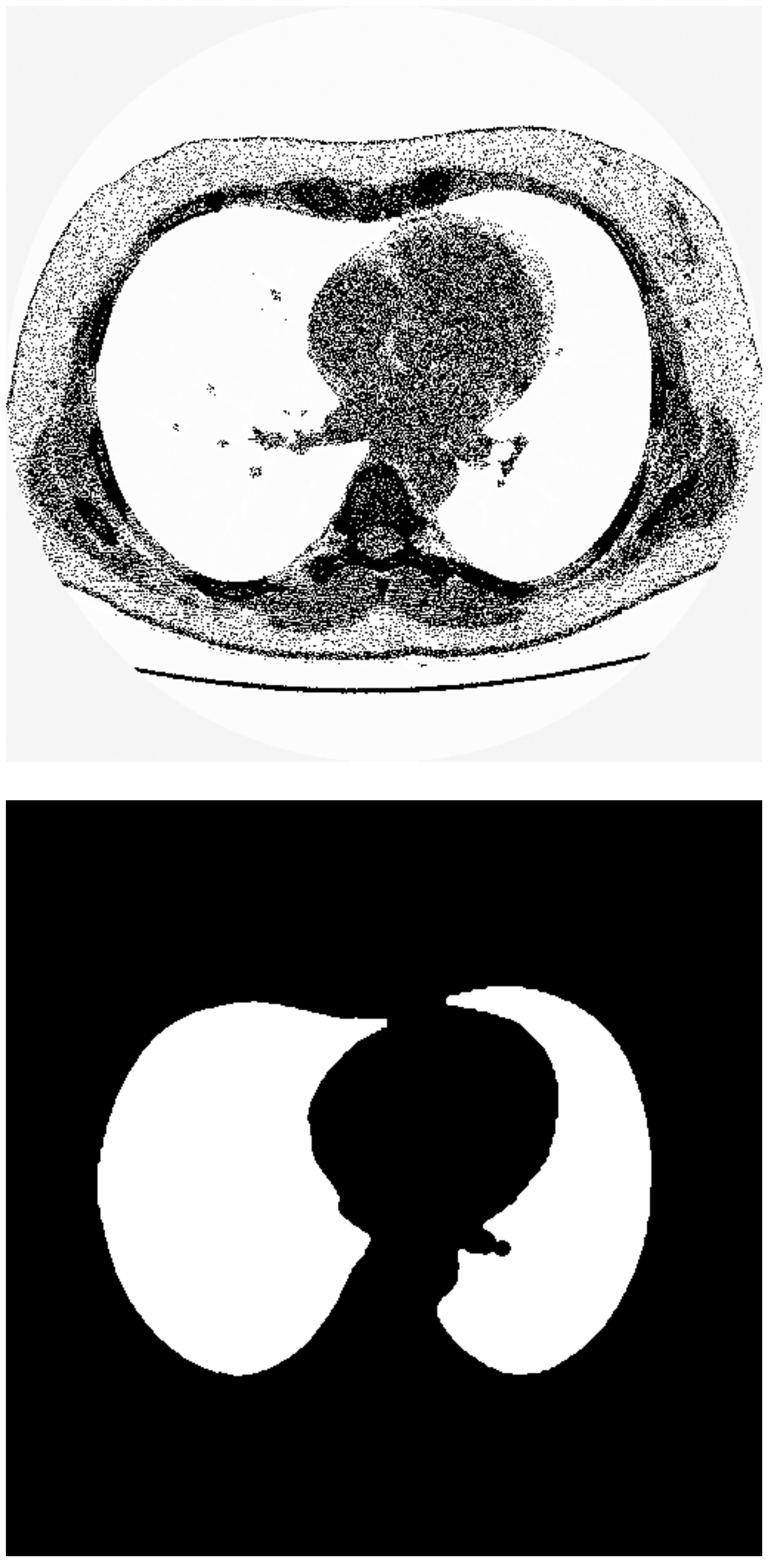}
         \caption{LUNA}
         \label{fig:luna}
     \end{subfigure}
     \hfill
     \begin{subfigure}[b]{0.16\textwidth}
         \centering
         \includegraphics[width=\textwidth]{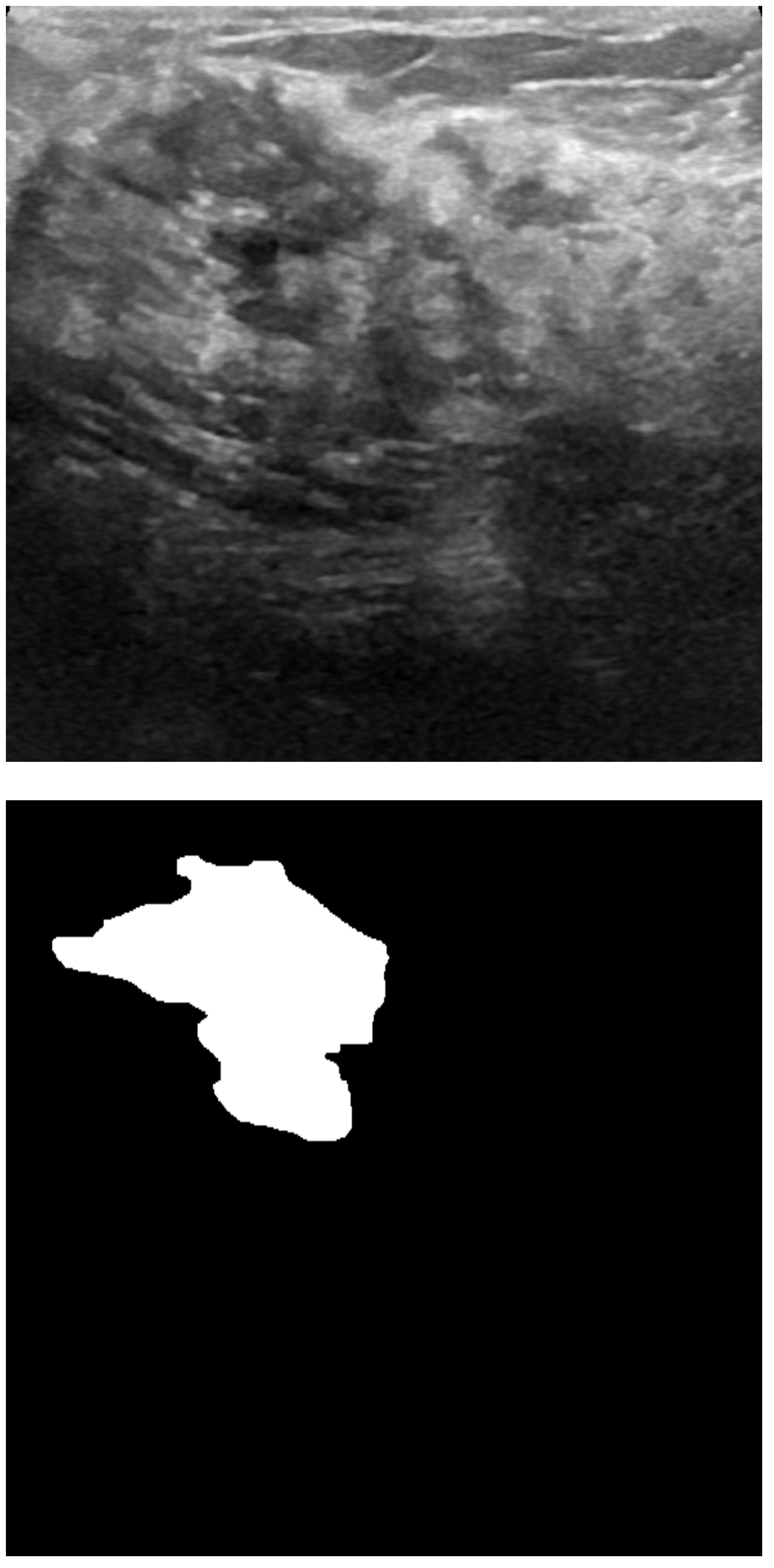}
         \caption{BUSI}
         \label{fig:busi}
     \end{subfigure}
     \hfill
     \begin{subfigure}[b]{0.16\textwidth}
         \centering
         \includegraphics[width=\textwidth]{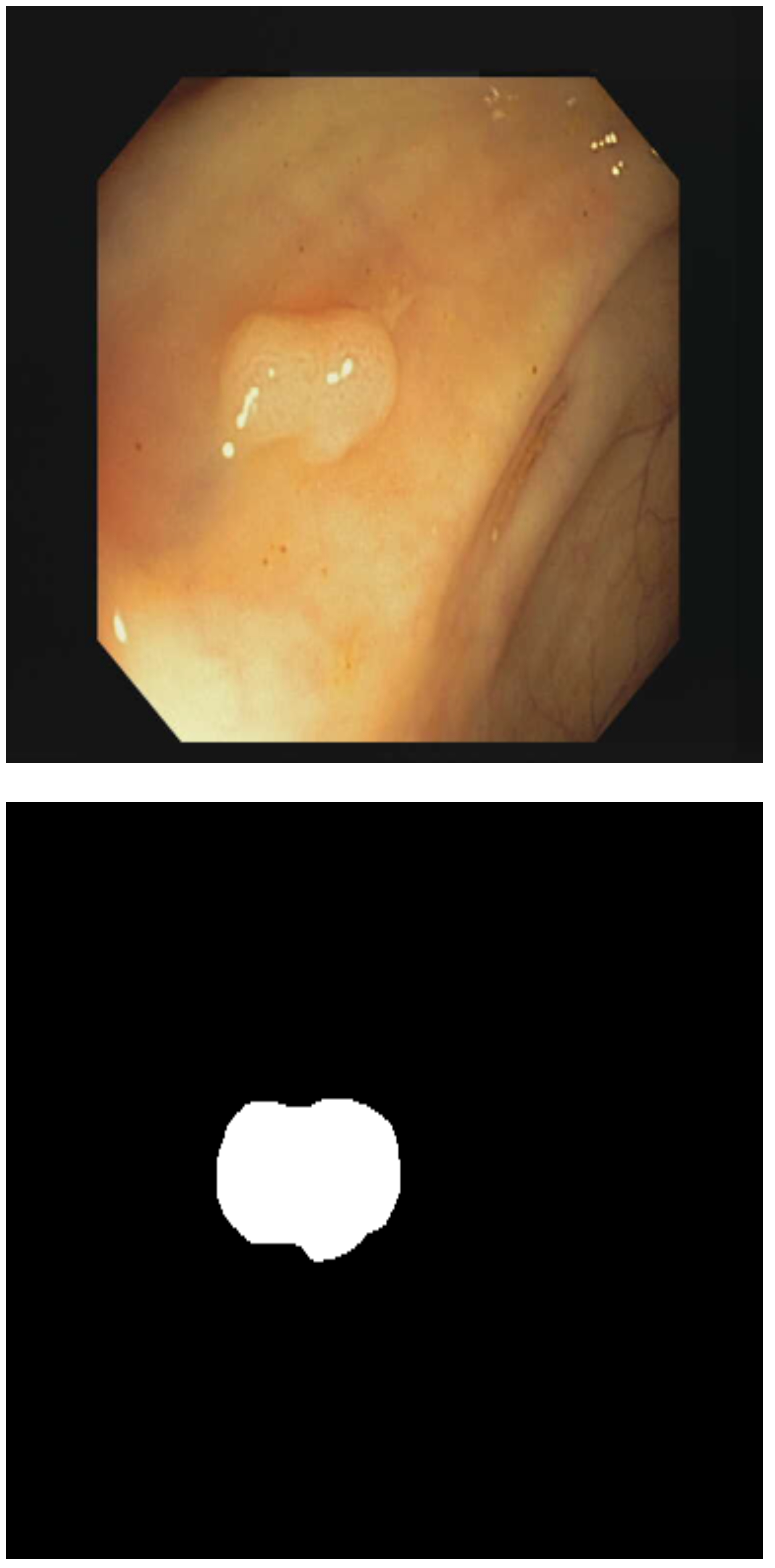}
         \caption{CVCclinicDB}
         \label{fig:cvcclinicdb}
     \end{subfigure}
     \hfill
     \begin{subfigure}[b]{0.16\textwidth}
         \centering
         \includegraphics[width=\textwidth]{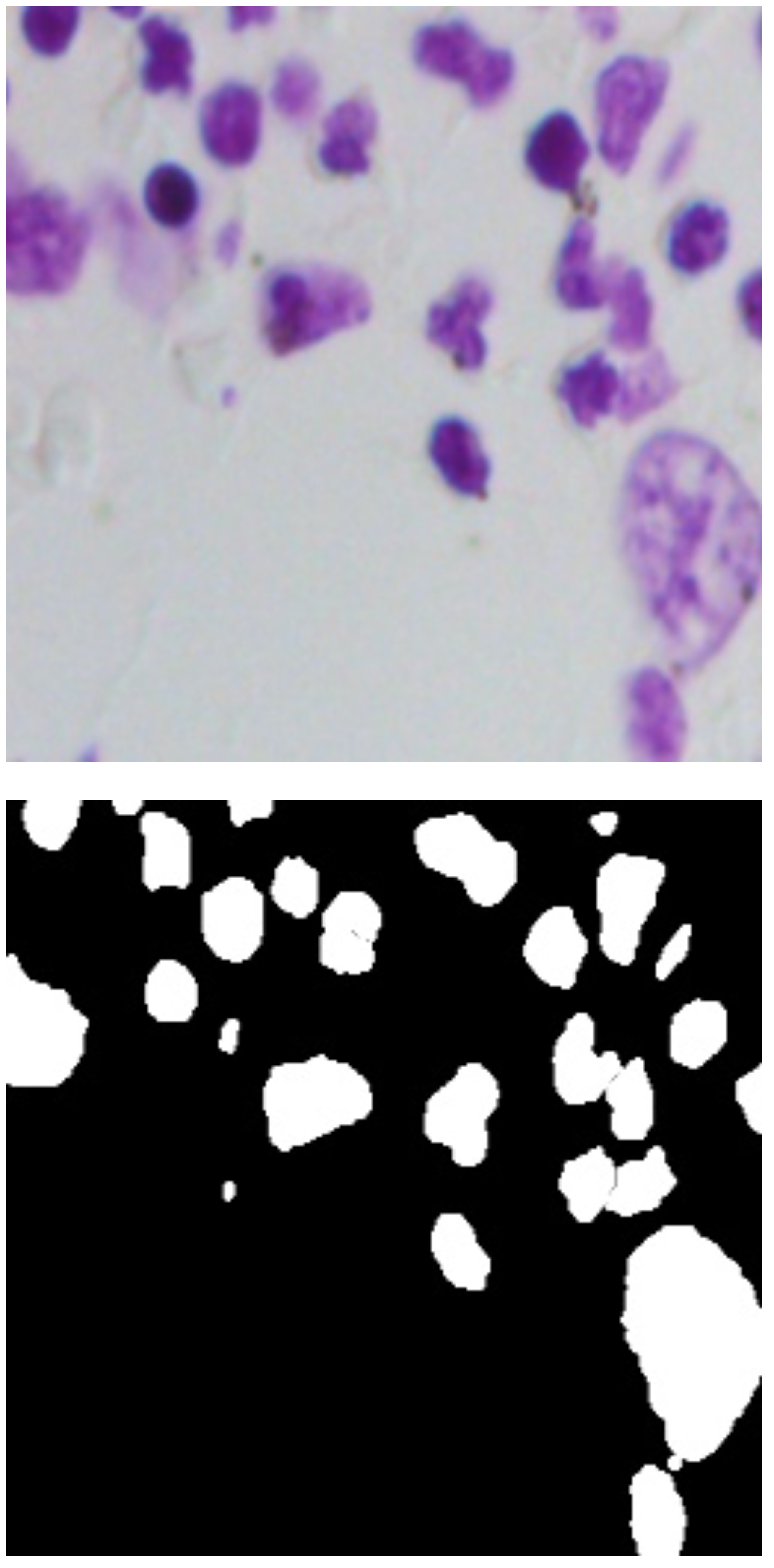}
         \caption{2018 DSB}
         \label{fig:dsb}
     \end{subfigure}
     \hfill
     \begin{subfigure}[b]{0.16\textwidth}
         \centering
         \includegraphics[width=\textwidth]{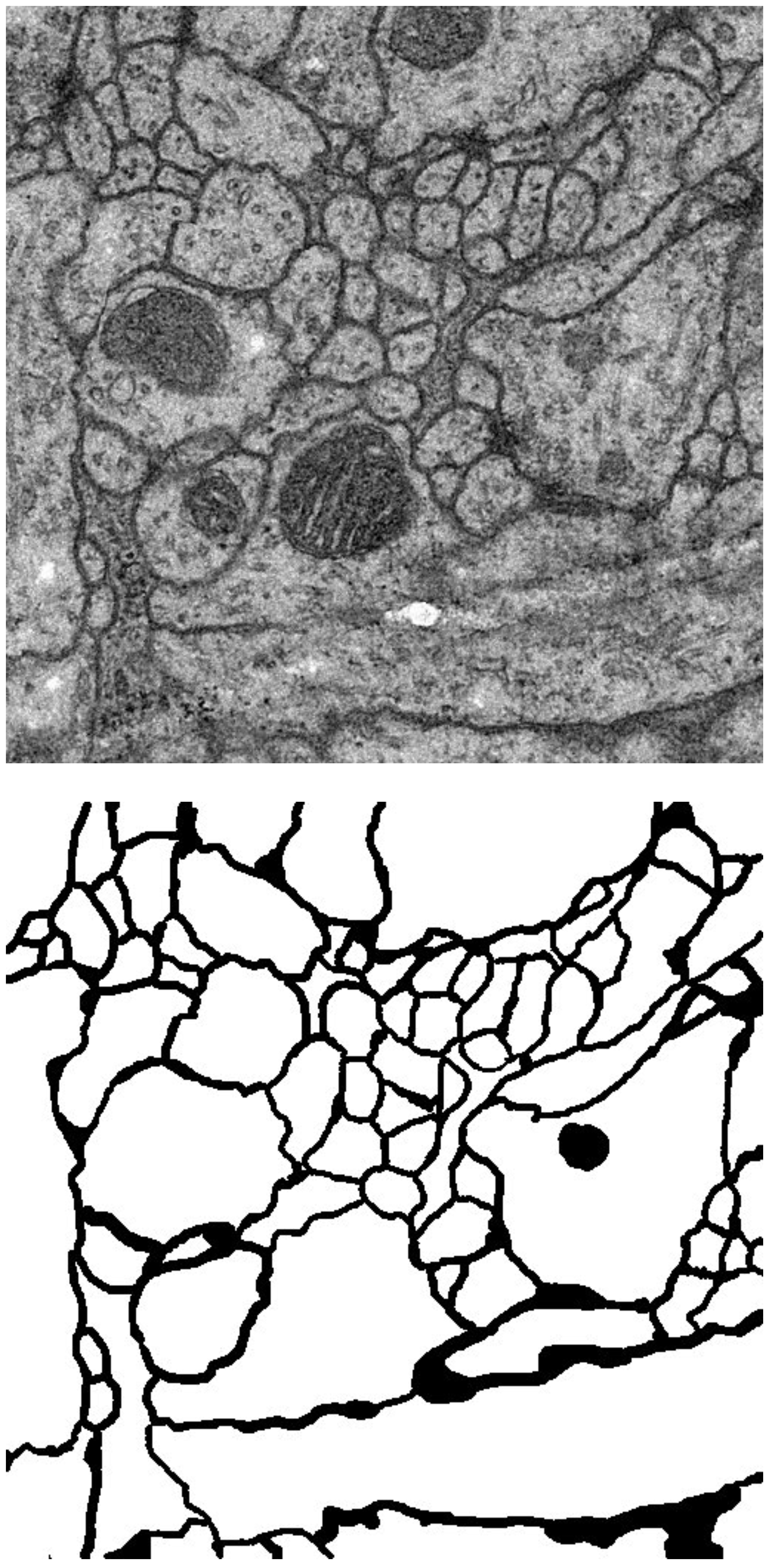}
         \caption{ISBI 2012}
         \label{fig:isbi}
     \end{subfigure}
        \caption{{\scriptsize Input images and their corresponding segmentation masks in the dataset. Sample images and their masks of DRIVE, LUNA, BUSI, CVCclinicDB, 2018 DSB and ISBI 2012 can be found in (a),(b),(c),(d),(e), and (f), respectively}}
        \label{fig:datasets}
\end{figure*}

This section briefly describes all the utilized datasets in this study. For the evaluation of the proposed model, we have utilized six datasets of different modalities, namely BUSI, CVCclinicDB, Drive, ISBI 2012, 2018 DSB, and LUNA. A representative image and the corresponding mask from each of the datasets are depicted in Fig. \ref{fig:datasets}.

\subsubsection{DRIVE}
A diabetic retinopathy screening program in the Netherlands provided the data used to create the Digital Retinal Images for Vessel Extraction (DRIVE) dataset facilitating retinal vessel segmentation as described in \cite{staal2004ridge}. A Canon CR5 non-mydriatic 3CCD camera with a 45-degree field of view (FOV) was used for image acquisition, and there is a total of 40, 8-bit per color channel images with a resolution of $768 \times 584$ pixels.

\subsubsection{Lung Segmentation}
Based on the computed tomography (CT) image modality, lung segmentation from CT images available in the Lung nodule analysis (LUNA) competition \cite{LUNADATASET}. This dataset contains 267 2D CT images with full annotations of the labeled lung images provided by experts in the medical sector. The size of each image is $512 \times 512$ pixels.

\subsubsection{Breast ultrasound image}
Utilizing LOGIQ E9 ultrasound system guided scanning, the breast ultrasound image (BUSI) dataset was created from images collected from 600 females aged between 25 to 75 years old ~\cite{ALDHABYANI2020104863}. The dataset contains seven hundred eighty images with an average image size of $500 \times 500$ pixels in three distinct categories: benign, normal, and malignant. The ground truth for each image was generated using Matlab.

\subsubsection{CVCclinicDB}
The CVCclinicDB dataset contains image frames extracted from colonoscopy videos, using Window Median Depth of Valley (WM-Dova) methodology as mentioned in ~\cite{bernal2015wm}. From a collection of twenty-nine video sequences, 612 still image frames were extracted for polyp detection. Each image is of the size $384 \times 288$, while the corresponding ground truth image  is presented as the segmentation mask of the polyps.

\subsubsection{2018 Data Science Bowl}
The 2018 data science bowl (DSB) dataset was created as a challenge for generic segmentation of nuclei of cells in a diverse set of stained two-dimensional (2D) microscopic images~\cite{caicedo2019nucleus}. The training set contains 670 images from both bright-stained and fluorescence modalities of microscopic images with sizes $256 \times256 \times 3$. In addition to the images captured under various lighting conditions, corresponding annotations (segmentation masks) for each image are also provided to be used as ground truth.

\subsubsection{ISBI 2012}
The ISBI 2012 dataset, introduced in ~\cite{10.1371/journal.pbio.1000502}, is comprised of transmission electron microscopy (TEM) images of \textit{Drosophila} larval brain for the purpose of analyzing the structural aspect of neural micro-circuitry. The training data is comprised of 30 TEM $512 \times 512$ serial section images of the first instar larval \textit{Dorsophila} brain using TrakEm2 ~\cite{10.1371/journal.pone.0038011} software. The corresponding labels of each image were produced by an expert neuro-anatomist for the purpose of segmentation.

\begin{table*}[!htb]
\footnotesize
  \caption{Overview of the datasets employed in our experiments.}
  \centering
  \begin{tabular*}{\textwidth}{@{\extracolsep{\fill}}lcccc@{\extracolsep{\fill}}}
   \toprule%
    Dataset & \#Training Samples & Modality & Original Shape & \#Samples after augmentation\\
   \midrule
    DRIVE & 20 & Retinal & $565\times 684$ & 368\\
    LUNA & 267 & Computed Tomography & $512\times512$ & 2992\\
    BUSI & 647 &  Ultrasound & $558\times462$ & 7080\\
    CVCclinicDB & 612 & Colonoscopy & $384\times288$ & 6856\\
    2018 DSB & 670 & Brightfield \& Fluorescence & $256\times256$ & 7504\\
    ISBI 2012 & 30 & Electron Microscopy & $512\times512$ & 528\\
   \botrule
  \end{tabular*}
  \label{tab:datasets}
\end{table*}

\subsubsection{Preprocessing and Data augmentation}
In our experiments, we used several augmentation techniques to ensure that over-fitting does not occur for a small number of samples present in the datasets. To ensure efficient, robust learning of the proposed model in five datasets, namely CVCclincDB, 2018 DSB, BUSI, and LUNA, we employed a total of thirteen data augmentation techniques, including two variations of random rotations, grid distortion, horizontal and vertical flips, transpose, a composition of vertical flip and random rotation, random brightness, random contrast, random brightness contrast, random gamma, hue-saturation contrast, and RGB shifting to increase the image variability during the training process. For the DRIVE and ISBI 2012 datasets, we employed a total of twenty-two data augmentation techniques, that includes the techniques mentioned above, as well as CLAHE, FancyPCA, and Gaussian noise injection. It should be noted that the original DoubleU-Net architecture employed a total of twenty-five augmentation types of a single image mask pair. The augmented RGB images after the data augmentation process were compressed to $256 \times 256$ to prepare them for fitting into the models. It should also be noted that the original images were also resized and incorporated into the training dataset.

\subsection{Training setup and experimental metrics}
In order to train the models, the augmented dataset was divided using an 80:10:10 ratio, i.e., 80\% of the images were used for composing the training dataset, 10\% for the testing, and the rest 10\% for the validation dataset. We initialize the pre-trained weights of EfficientNetB7 architecture, and the batch size was set at 4. The learning rate starts from 0.0001, and the learning rate is reduced by a factor of 0.1, with a patience of 10. We fed 2D images of size $256 \times 256$ as input for the proposed network. Our system was implemented using Tesla P100-PCIE GPU with 16 GB RAM and a Tensorflow backend. The total number of trainable parameters of the proposed model is 22.4 million. We incorporated a hybrid loss function by adding the Binary Cross-Entropy loss ($Loss_{BCE}$) and Dice loss ($Loss_{Dice}$)~\cite{Sudre2017}, offering smooth gradient flow and handling of the class imbalance problems~\cite{bose2022dense}. The hybrid loss function can be defined as:
\begin{equation}
    Loss_{Hybrid} =  Loss_{BCE} + Loss_{Dice}
    \label{eq:hybrid_loss}
\end{equation}
\begin{equation}
\begin{split}
    Loss_{BCE} &= -\sum_{c=1}^My_{o,c}\log(p_{o,c}) \\
               &= -{(y\log(p) + (1 - y)\log(1 - p))}
\end{split}
\label{eq:bce_loss}
\end{equation}

The $Loss_{BCE}$ specified in equation \ref{eq:bce_loss} can be defined in terms of number of classes $M$, the natural $log$, binary indicator (0 or 1) $y$, class label $c$ the correct classification for observation $o$, and $p$ is the predicted probability observation $o$ is of class $c$.

\begin{equation}
    Loss_{Dice} = 1 - \frac{2\sum_{i=1}^{N}p_{i}g_{i}+\epsilon}{\sum_{i=1}^{N}p_{i}^{2}+\sum_{i=1}^{N}g_{i}^{2}+\epsilon}
    \label{eq:dice_loss}
\end{equation}

The dice coefficient between the prediction samples $p$ and the mask $g$ can be defined as given in equation \ref{eq:dice_loss}. Here, $\epsilon$ is a constant added to avoid the divide by zero error.

\subsubsection{Precision and Recall}
 True positive (TP) outcomes are the number of samples that were correctly classified as the mask, and false positives (FP) are the number of samples that were falsely predicted as part of the mask region. On the other hand, the true negatives (TN) are the number of samples that are correctly classified as not present inside the mask region, and the false negatives (FN) are the pixels that are falsely classified as not present inside the masked region. Thus, we can now calculate the precision and recall from the confusion matrix as follows:

\begin{equation}
    \textnormal{Precision} = \frac{TP}{TP+FP}
\end{equation}

\begin{equation}
    \textnormal{Recall} = \frac{TP}{TP+FN}
\end{equation}

\subsubsection{DICE similarity coefficient}
The DICE coefficients, as coined by \cite{dice1945measures}, are used widely for image segmentation purpose, and it has been used in the case of both 2D and 3D image segmentation tasks. The dice coefficients required for image segmentation can be constructed from a contingency table \cite{Zou2004} of four possible outcomes as represented in the probabilities of segmentation results from an image. DSC can be generalized using the definitions of true positives (TP), false positives (FP), and false negatives (FN) as:
\begin{equation}
        \textnormal{DSC} = \frac{2TP}{2TP+FP+FN}
\end{equation}

The dice coefficient measures how much the area of interest of two images has overlapped. DSC values have a range of [0,1]. The higher the DSC value is, the better segmentation is achieved from the prediction. 

\subsubsection{Intersection-Over-Union (IoU)} 
 
Along with DSC, mean Intersection-Over-Union (mIoU) can be used to calculate the prediction similarity with ground truth. IoU values have a range of [0,1]. The higher value of IoU, means there is a better similarity between prediction and ground truth. IoU can be defined in terms of the common confusion metrics as follows:
\begin{equation}
    \textnormal{IoU} = \frac{TP}{TP+FP+FN}
\end{equation}

\subsection{Evaluation of the segmentation results}

This section provides the quantitative and qualitative analysis result analysis of the proposed DoubleU-NetPlus method with other SOTA methods.

\begin{figure*}[!htb]
     \centering
     \begin{subfigure}[b]{0.64\textwidth}
         \centering
         \includegraphics[width=\textwidth]{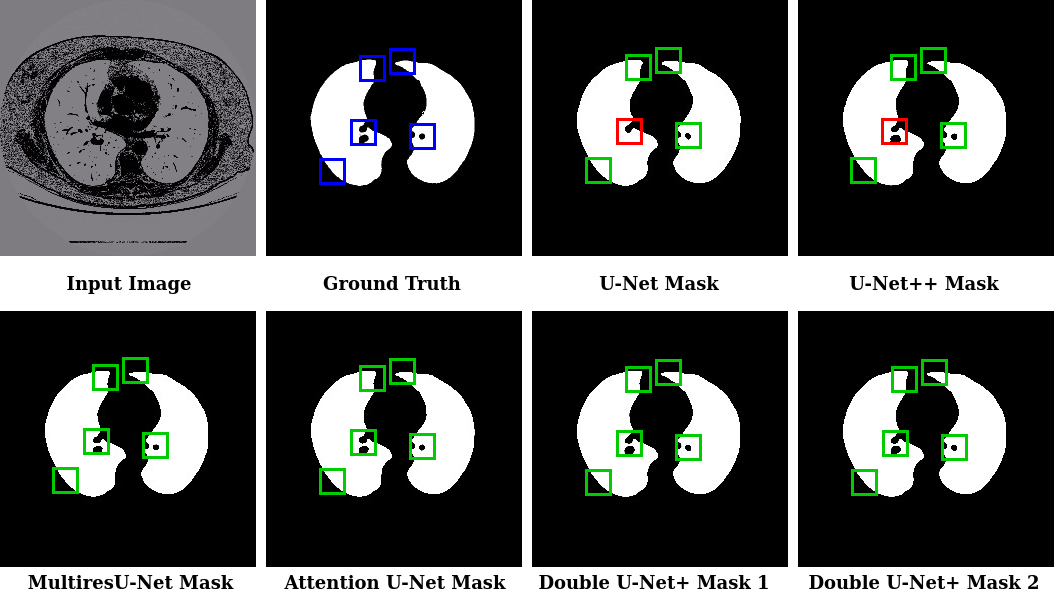}
         \caption{LUNA}
         \label{fig:luna-compare}
     \end{subfigure}
     \hfill
     \begin{subfigure}[b]{0.64\textwidth}
         \centering
         \includegraphics[width=\textwidth]{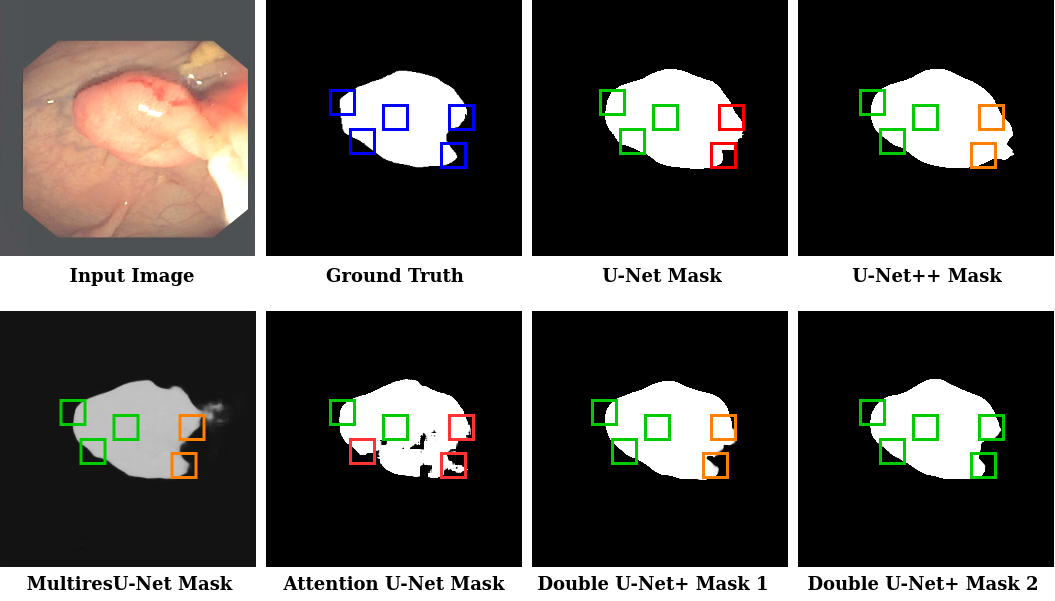}
         \caption{CVCclinicDB}
         \label{fig:cvc-compare}
     \end{subfigure}
     \hfill
     \begin{subfigure}[b]{0.64\textwidth}
         \centering
         \includegraphics[width=\textwidth]{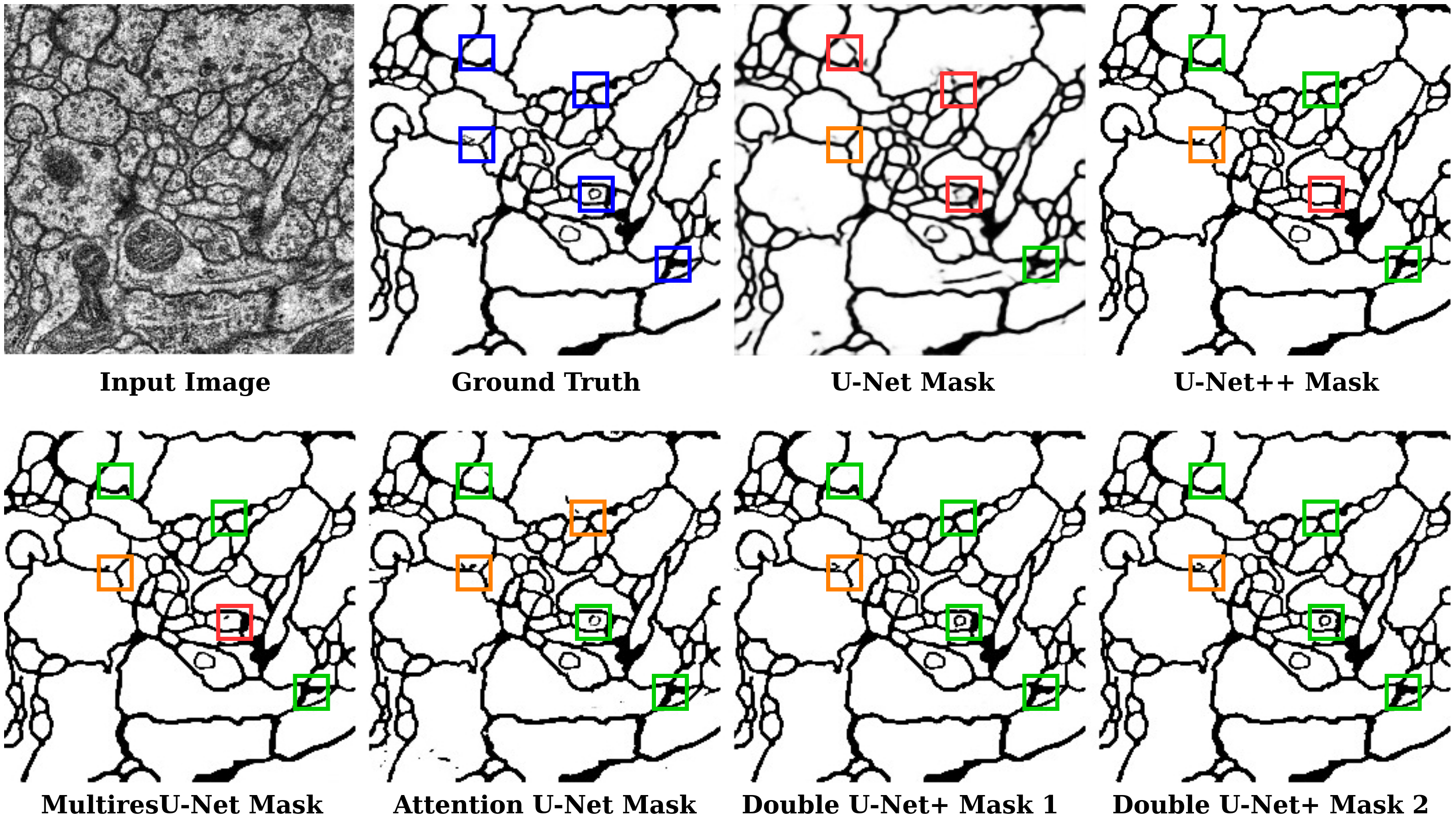}
         \caption{ISBI 2012}
         \label{fig:isbi-compare}
     \end{subfigure}
     \footnotesize
     \caption{\footnotesize Visual comparative analysis among different segmentation methods. First row (left to right): input image, ground truth, U-Net output, U-Net++ output. Second row (left to right): MultiResU-Net output, Attention U-Net output, and results of the DoubleU-NetPlus network (Mask 1 and 2) on the LUNA dataset. A similar pattern is followed in rows three to six for CVCclinicDB and ISBI 2012 datasets. Blue, red, yellow, and green boxes denote exemplary ROI, unsatisfactory, moderate, and good results.}
     \label{fig:data-compare}
\end{figure*}

\begin{figure*}[!htb]
     \centering
     \begin{subfigure}[b]{0.64\textwidth}
         \centering
         \includegraphics[width=\textwidth]{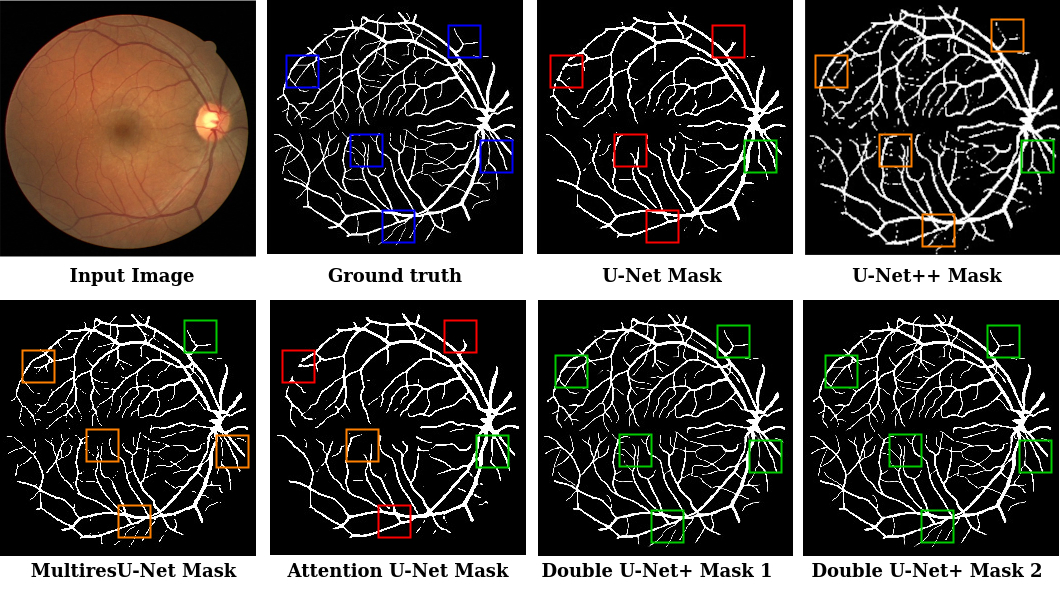}
         \caption{DRIVE}
         \label{fig:drive-compare}
     \end{subfigure}
     \hfill
     \begin{subfigure}[b]{0.64\textwidth}
         \centering
         \includegraphics[width=\textwidth]{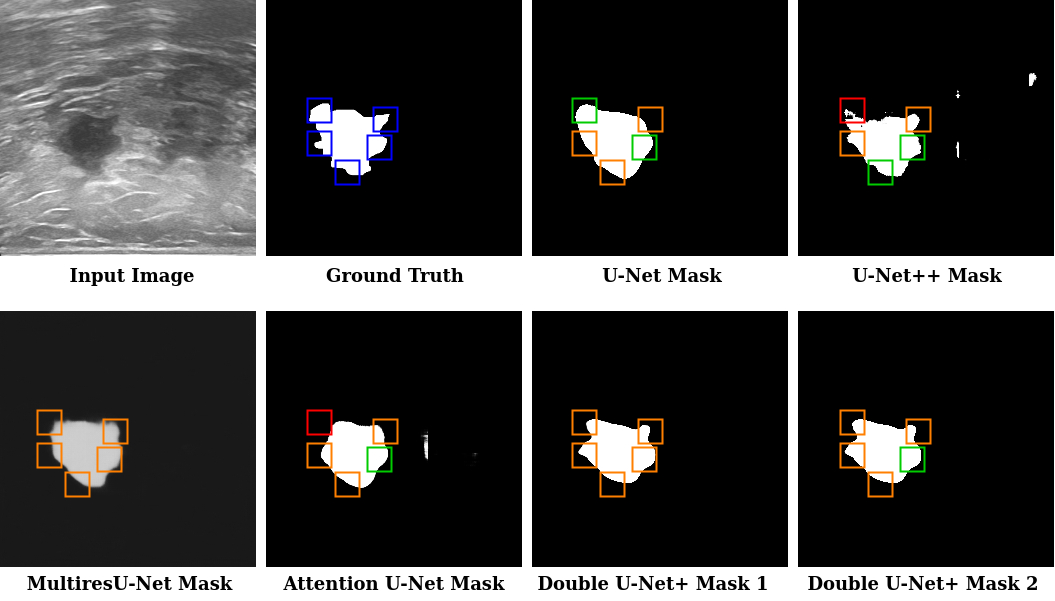}
         \caption{BUSI}
         \label{fig:busi-compare}
     \end{subfigure}
     \hfill
     \begin{subfigure}[b]{0.64\textwidth}
         \centering
         \includegraphics[width=\textwidth]{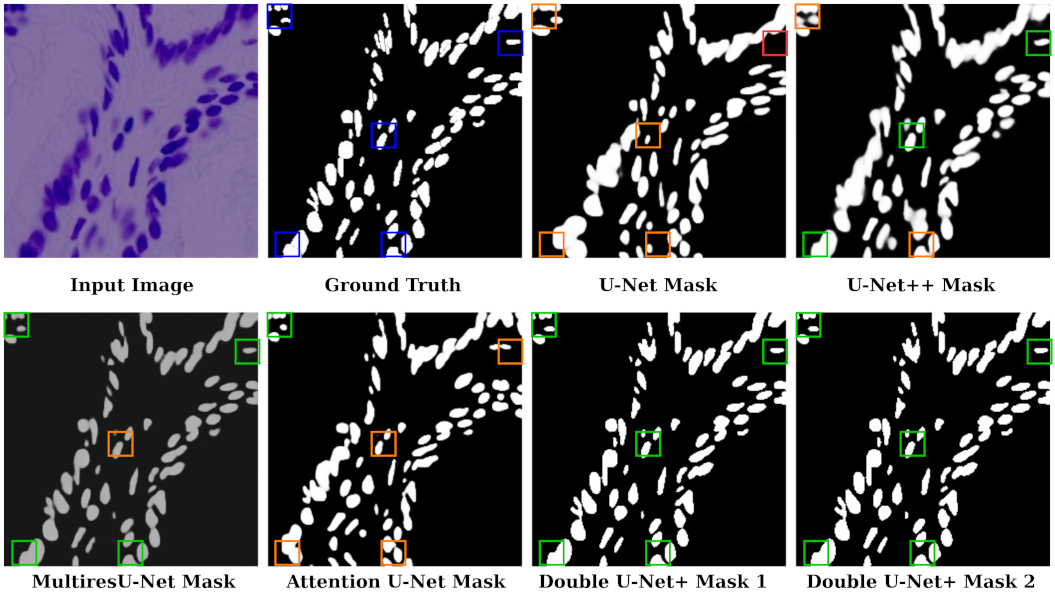}
         \caption{2018 DSB}
         \label{fig:DSB-compare}
     \end{subfigure}
     \footnotesize
    \caption{\footnotesize Visual comparative analysis among different segmentation methods. First row (left to right): input image, ground truth, U-Net output, U-Net++ output. Second row (left to right): MultiResU-Net output, Attention U-Net output, and results of the DoubleU-NetPlus network (Mask 1 and 2) on the DRIVE dataset. A similar pattern is followed in rows three to six for BUSI and 2018 DSB datasets. Blue, red, yellow, and green boxes denote exemplary ROI, unsatisfactory, moderate, and good results.}
     \label{fig:data2-compare}
\end{figure*}

\begin{table*}[!htb]
  \footnotesize
  \caption{Comparisons of the segmentation result for the proposed and conventional methods in all the employed datasets. The numbers in bold indicate the best performance of the corresponding indicators."-" denotes that there is no backbone used in the network.}
  \centering
  \resizebox{0.839\textwidth}{!}{
  \begin{tabular*}{\textwidth}{@{\extracolsep{\fill}}lcccccc@{\extracolsep{\fill}}}
   \toprule
    Dataset & Model & Backbone & Precision & Recall & DICE & mIoU\\

    \midrule
    DRIVE & U-Net~\cite{U-net} & -& 94.46 & 88.32 & 77.19 & 62.94\\
    & U-Net++~\cite{Unet++} & -& 97.31 & 90.59 & 82.24 & 70.44\\
    & Attention UNet~\cite{oktay2018attention} &- & 97.55 & 92.49 & 80.51 & 67.52\\
    & MultiResU-Net~\cite{MultiResUNet} &  & 97.20 & 88.12 & 83.08 & 71.17\\
    & FA-Net~\cite{Fanet} & - & 81.89 & 82.15 & 81.83 & 69.27\\
    & ConvUNeXt~\cite{ConvUNeXt} & ConvNeXt & - & - & 82.30 & \textbf{82.60}\\
    & LCP-Net~\cite{LCP-Net} & - & 89.69 & 78.72 & - & 82.14\\
    & Res2UNet\cite{Res2Unet} & -& - & - & 81.86 & 69.26\\
    & LadderNet~\cite{LadderNet} & -& - &  78.56 & 82.02 & -\\
    & DoubleU-Net~\cite{Doubleu-net} & VGG-19 & 94.22 & 91.63 & 83.22 & 70.82\\
    & IterNet++~\cite{IterNet++} & -& - &  83.99 & 83.13 & 71.15\\
    & Sharp U-Net~\cite{sharpunet} & ResNet-50 & - & - & 82.03 & 71.43\\
    & CE-Net~\cite{Ce-net} & -& - & 83.09 & 81.35 & 68.54\\
    & R2U-Net~\cite{r2u-net} & -& - &  - & 81.24 & 68.38\\
    & SUD-GAN~\cite{SUD-GAN} & -& 88.21 &  83.40 & - & -\\
    & \textbf{DoubleU-NetPlus} & EfficientNetB7&  \textbf{98.05} &  \textbf{96.48} & \textbf{85.17} & 73.92\\
    
    \midrule
    LUNA & U-Net~\cite{U-net}&- & 97.42 & 95.36 & 95.11 & 91.30\\
    & U-Net++~\cite{Unet++} & -& 99.29 & 98.48 & 94.07 & 93.61\\
    & Attention U-Net~\cite{oktay2018attention} &- & 99.25 & 96.96 & 93.60 & 93.09\\
    & MultiResU-Net~\cite{MultiResUNet} &  & 99.33 & 97.63 & 92.14 & 90.07\\
    & CE-Net~\cite{Ce-net} & -& - & 98.00 & - & 96.20\\
    & DCU-Net~\cite{DCU-Net} & -& 97.31 & 96.99 & 97.33 & 96.17\\
    & DoubleU-Net~\cite{Doubleu-net} & VGG-19 & 95.68 & 89.20 & 97.74 & 95.11\\
    & FBUNet~\cite{FBUNet} & -& 98.33 & 98.67 & 98.50 & 97.06 \\
    & EANet~\cite{EANet} & VGG-19 & 98.66 &  98.77 & 98.65 & -\\
    & Sharp U-Net~\cite{sharpunet} & ResNet-50 & - & 98.77 & 97.25 & 95.22\\
    & FA-Net~\cite{Fanet} & - & 83.16 & 84.14 & 95.44 & 93.70\\
    & Bose et al.~\cite{bose2022dense} & - & - &  97.90 & - & 97.00\\
    & \textbf{DoubleU-NetPlus} & EfficientNetB7 & \textbf{99.57} &  \textbf{98.82} & \textbf{99.34} & \textbf{98.93}\\
    
    \midrule
    BUSI & U-Net~\cite{U-net} & - & 94.04 & 84.67 & 80.44 & 63.86\\
    & U-Net++~\cite{Unet++} &- & 93.81 & 85.19 &  81.29 &  65.55\\
    & Attention U-Net~\cite{oktay2018attention} & - & 91.06 & 77.31 & 81.34 & 68.10\\
    & MultiResU-Net~\cite{MultiResUNet} &- & 96.77 & 85.15 & 78.76 & 63.60\\
    & FA-Net~\cite{Fanet} & - & 93.10 & 88.73 & 90.63 & 71.95\\
    & CE-Net~\cite{Ce-net} &- & - & - & 80.69 & 67.63\\
    & MCRNet~\cite{MCRNet} & ResNet-34 & - &  81.92 & 82.31 & 69.94\\
    & RCA-IUnet~\cite{punn2022rca} & -& 94.01 &  89.11 & 91.40 & \textbf{89.95}\\
    & DoubleU-Net~\cite{Doubleu-net} & VGG-19 & 93.11 & 88.85 & 89.93 & 71.49\\
    & Sharp U-Net~\cite{sharpunet} & ResNet-50 & - & 92.42 & 90.82 & 74.20\\
    & CFPNet~\cite{CFPNet-M} & -& - &  - & 80.92 & 67.95\\
    & GC-Net~\cite{xue2021global} & -& 86.5 &  - & 82.10 & 73.80\\
    & \textbf{DoubleU-NetPlus} & EfficientNetB7 & \textbf{96.90} &  \textbf{92.47} & \textbf{94.30} & 84.71\\
    
    \midrule
    CVCclinicDB & U-Net~\cite{U-net} & -& 95.21 & 89.35 & 88.11 & 83.52\\
    & U-Net++~\cite{Unet++} & -& 92.11 &  82.23 & 87.40 & 77.27\\
    & Attention U-Net~\cite{oktay2018attention} & - & 94.20 & 92.00 & 93.88 & 88.87\\
    & MultiResU-Net~\cite{MultiResUNet} &- & 95.96 & 92.59 & 94.10 & 89.96\\
    & DoubleU-Net~\cite{Doubleu-net} & VGG-19 & 95.92 &  84.57 & 92.39 & 86.11\\
    & CRF-EfficientUNet~\cite{crf} & EfficientNetB7 & 96.83 & \textbf{97.94} & 95.12 & 91.85\\
    & MSRF-Net~\cite{Msrf-net} & -& 94.27 & 95.67 & 94.20 & 90.43\\
    & Sharp U-Net~\cite{sharpunet} & ResNet-50 & - & - & 90.05 & 83.98\\
    & AMNet~\cite{song2022attention} & Res2Net & - &  - & 93.60 & 88.80\\
    & FA-Net~\cite{Fanet} & - & 94.01 & 93.39 & 93.55 & 89.37\\
    & Focus U-Net~\cite{Focus} & - & 93.00 & 95.60 & 94.10 & 89.30\\
    & Ds-transunet~\cite{Ds-transunet} & Transformer & 93.69 &  95.00 & 93.80 & 89.10\\
    & Swin U-Net~\cite{Swin-unet} & Transformer & - &  - & 92.30 & 87.50\\
    & Polyp-PVT~\cite{Polyp-pvt} & Transformer & - &  - & 93.70 & 88.90\\
    & \textbf{DoubleU-NetPlus} & EfficientNetB7 & \textbf{97.96} &  93.87 & \textbf{96.40} & \textbf{95.12}\\
    
    \midrule
    2018 DSB & U-Net~\cite{U-net} & ResNet-101 & 96.84 &  78.55 & 92.09 & 85.60 \\
    & U-Net++~\cite{Unet++} &- & 91.99 &  79.41 & 85.81 & 85.52\\
    & Attention U-Net~\cite{oktay2018attention} &- & 98.09 & 80.07 & 93.21 & 87.61\\
    & MultiResU-Net~\cite{MultiResUNet} & - & 98.45 & 83.77 & 87.55 & 83.51\\
    & DoubleU-Net~\cite{Doubleu-net} & VGG-19 & 94.96 & 94.07 & 92.33 & 88.07\\
    & Sharp U-Net~\cite{sharpunet} & ResNet-50 & - & - & 95.40 & 89.60\\
    & FA-Net~\cite{Fanet} & - & 91.94 & 92.22 & 91.76 & 85.69\\
    & MSRF-Net~\cite{Msrf-net} & -& 90.22 &  94.02 & 92.24 & 85.34\\
    & Poudel and Lee~\cite{poudel2021deep} &  EfficientNetB3 & - &  - & 90.07 & \textbf{90.97}\\
    & \textbf{DoubleU-NetPlus} & EfficientNetB7 & \textbf{98.82} &  \textbf{95.64} & \textbf{95.76} & 90.29\\
    
    \midrule
    ISBI 2012 & U-Net~\cite{U-net} & -& 99.53 & 81.58 & 94.30 & 89.38\\
    & U-Net++~\cite{Unet++} &- & 94.61 &  95.56 & 93.94 & 88.60\\
    & Attention U-Net~\cite{oktay2018attention} & - & 99.70 & 82.20 & 96.80 & 93.81\\
    & MultiResU-Net~\cite{MultiResUNet} & -& 99.72 & 82.50 & 96.13 & 92.90\\
    & FBUNet~\cite{FBUNet} & -& 93.73 & 94.19 & 93.96 & 88.62\\
    & LCP-Net~\cite{LCP-Net} &- & 89.69 & \textbf{98.95} &  \textbf{98.12} & 82.14\\
    & DoubleU-Net~\cite{Doubleu-net} & VGG-19 & 95.85 & 92.36 & 93.60 & 89.87\\
    & Sharp U-Net~\cite{sharpunet} & ResNet-50 & - & - & 93.52 & 91.21\\
    & FA-Net~\cite{Fanet} & - & 95.29 & 95.68 & 95.47 & 91.34\\
    & \textbf{DoubleU-NetPlus} & EfficientNetB7 &  \textbf{99.75} &  88.62 & 97.10 &  \textbf{94.38}\\
   \botrule
  \end{tabular*}
  }
  \label{tab:comparison}
\end{table*}

\subsubsection{Quantitative result analysis}
Here, we report the quantitative results on six various modalities of medical image datasets and compare them to other SOTA approaches to ensure that the proposed model surpasses the performance or performs at par with other SOTA methods (on the same train-test split ratio and similar types of data augmentation methods). It is important to note that, in order to provide a fair comparison, the evaluation metrics are provided only for the approaches that prioritize segmentation performance over computational efficiency. The performance of the model on all the utilized datasets can be observed in Table~\ref{tab:comparison}.

\textbf{\textit{Results on DRIVE:}} A comparison with well-established segmentation architectures with different backbones demonstrates that our proposed method outperforms the SOTA architectures. With a Dice score of 85.17\%, mIoU of 73.92\%, precision of 98.05\%, and recall of 96.48\% (see Table~\ref{tab:comparison}), the DoubleU-NetPlus architecture significantly surpasses all SOTA architectures on the DRIVE dataset. While outperforming U-Net and most of its variants, it can also be observed that DoubleU-NetPlus exceeds the performance of the recently proposed ConvNeXt~\cite{ConvNeXt} encoder backbone-based ConvUNeXt~\cite{ConvUNeXt} architecture by a Dice score of 2.87\% though ConvUNeXt reports the highest mIoU value of 82.60\%. Compared to FANet~\cite{Fanet}, the model achieves an increase of 4.65\% in the mIoU metric with a lesser number of augmented images during training.

\textbf{\textit{Results on LUNA:}} In the LUNA dataset, the DouleU-NetPlus network achieves SOTA segmentation results of 99.34\% on Dice, 98.93\% on mIoU, 99.57\% on precision, and 98.82\% on recall metrics, respectively (see Table~\ref{tab:comparison}). The results outperform U-Net~\cite{U-net}, U-Net++~\cite{Unet++}, VGG-19 encoder-based EANet~\cite{EANet}, and ResNet-50 based Sharp U-Net~\cite{sharpunet} architectures in the Dice metric by a margin of 4.23\%, 5.27\%, 0.69\%, and 2.09\%. DoubleU-NetPlus also has the best balance on both the precision-recall and Dice-mIoU pairs.

\textbf{\textit{Results on BUSI:}} In the BUSI dataset, the DoubleU-NetPlus achieves significantly improved results compared to all the SOTA architectures. It achieves a precision of 96.90\% and a recall of 92.47\%. The model achieves a significantly improved Dice value of 94.30\% which is 13.01\% and 15.54\% better compared to the UNet++~\cite{Unet++}, and MultiResUNet~\cite{MultiResUNet} architectures respectively (see Table~\ref{tab:comparison}). Although the highest mIoU is achieved by RCA-IUnet~\cite{punn2022rca} with 89.95\%, compared to DoubleU-NetPlus's 84.71\%.

\textbf{\textit{Results on CVCclinicDB:}} Table~\ref{tab:comparison} demonstrates that in the CVCclinicDB dataset, DoubleU-NetPlus produces a Dice score of 96.40\%, mIoU of 95.12\%, precision of 97.96\%, and a recall value of 93.87\% with an improvement of 4.01\% in Dice with respect to SOTA DoubleU-Net architecture. Our model achieves the best trade-off between Dice and mIoU metrics compared with the SOTA architectures resulting in the highest mIoU metric value of 95.12\%, surpassing the dual Swin Transformer-based Ds-transunet~\cite{Ds-transunet} model by 6.02\% in the mIoU metric.

\textbf{\textit{Results on 2018 DSB:}} DoubleU-NetPlus obtains significantly improved precision value of 98.82\%, Dice of 95.76\%, and mIoU of 90.29\% which are much improved results compared to U-Net~\cite{U-net}, U-Net++~\cite{Unet++}, and DoubleU-Net~\cite{Doubleu-net} (see Table~\ref{tab:comparison}). It also achieves the best trade-off between Dice-mIoU compared to other SOTA architectures. Though Sharp U-Net~\cite{sharpunet} reports the highest Dice value of 95.40\%, in terms of mIou, DoubleU-NetPlus generates better results. Poudel and Lee~\cite{poudel2021deep} report the highest mIoU of 90.97\%; however, DoubleU-NetPlus outperforms their architecture by 5.69\% in the Dice metric.

\textbf{\textit{Results on ISBI 2012:}} In the ISBI 2012 dataset, DoubleU-NetPlus achieves 99.75\% in precision, 88.62\% in the recall, 97.10\% in Dice, and 94.38\% in mIoU metric, which are significantly improved results compared to the U-Net~\cite{U-net}, U-Net++~\cite{Unet++}, and MultiResU-Net~\cite{MultiResUNet} architectures. Especially in the mIoU metric, the proposed model obtains an increase of 5.00\%, 5.78\%, 0.57\%, and 1.48\% compared to U-Net, U-Net++, Attention U-Net~\cite{oktay2018attention}, and MultiResU-Net architectures respectively (see Table~\ref{tab:comparison}). The highest Dice value of 98.12\% is reported in LCP-Net~\cite{LCP-Net}.

The results of the DoubleU-NetPlus model show that the proposed model greatly improves the performance of MIS tasks in diverse modalities of colonoscopy, fluorescence, electron microscopy, CT, retinal, and ultrasound.

\subsubsection{Qualitative result analysis} 
The results that were obtained from the experiments on six datasets of diverse modalities were evaluated critically on visual qualitative criteria to ensure proper segmentation performance. Specifically, we illustrate the predictions of U-Net, U-Net++, Attention U-Net, MultiResUnet, and our proposed DoubleU-NetPlus segmentation architectures, which were also applied in the quantitative comparisons too. The visual comparisons of the mentioned architectures with the proposed DoubleU-NetPlus, as demonstrated in Fig.~\ref{fig:luna-compare}, ~\ref{fig:cvc-compare}, ~\ref{fig:isbi-compare}, and ~\ref {fig:drive-compare}, ~\ref{fig:busi-compare}, and ~\ref {fig:DSB-compare}, shows that the segmentation map of the DoubleU-NetPlus network achieves better semantic segmentation performance in every datasets. On visual inspection, it is clear that there are several instances where the proposed network outperforms SOTA architectures such as the U-Net, U-Net++, Attention U-Net, and MultiResUnet.

\begin{table*}[!htb]
  \footnotesize
  \caption{\scriptsize Ablation experiments that analyze the contributions of the different modules on the utilized datasets. The numbers in bold indicate the best performance of the corresponding indicators.}
  \begin{tabular*}{\textwidth}{@{\extracolsep{\fill}}llcccc@{\extracolsep{\fill}}}
   \toprule
    Dataset & Model & Precision & Recall & DICE & mIoU\\
    & &($\pm 0.5\%$) &($\pm 0.5\%$) &($\pm 0.5\%$) &($\pm 0.5\%$) \\
   \midrule
    Drive & Baseline & 98.42 & 96.36 & 97.90 & 96.64\\
    & DoubleU-NetPlus w/o MKRC & 97.51 & 95.81 & 83.27 & 71.53\\
    & DoubleU-NetPlus w/o TAM & 97.53 & 95.67 & 83.65 & 72.12\\
    & DoubleU-NetPlus w/o TAG & 97.55  & 96.98 &  83.71 & 72.09 \\
    & DoubleU-NetPlus w/o (TAM \& MKRC) & 96.52 & 97.08 & 82.29 & 70.00\\
    & DoubleU-NetPlus w/o (TAM \& MKRC \& TAG) & 97.19 &  96.11 & 82.30 & 70.24\\
    & DoubleU-NetPlus &  \textbf{98.05} &  \textbf{96.48} & \textbf{85.17} & 73.92\\

    \midrule
    LUNA & Baseline & 98.42 & 96.36 & 97.90 & 96.64\\
    & DoubleU-NetPlus w/o MKRC & 98.04 & 98.10 & 98.40 & 97.07\\
    & DoubleU-NetPlus w/o TAM & 98.79 & 98.19 & 98.61 & 97.53\\
    & DoubleU-NetPlus w/o TAG & 98.83 &  97.29 & 98.10 & 96.90\\
    & DoubleU-NetPlus w/o (TAM \& MKRC) & 98.10 & 98.25 & 98.11 & 96.83\\
    & DoubleU-NetPlus w/o (TAM \& MKRC \& TAG) & 97.91 &  98.05 & 97.92 & 96.77\\
    & DoubleU-NetPlus & \textbf{99.57} &  \textbf{98.82} & \textbf{99.34} & \textbf{98.93}\\
    
   \midrule
    BUSI & Baseline & 94.04 & 84.67 & 88.47 & 80.67\\
    & DoubleU-NetPlus w/o MKRC & 96.11 & 90.48 & 93.74 & 79.32\\
    & DoubleU-NetPlus w/o TAM & \textbf{97.36} & 91.78 & 93.84 & 75.11\\
    & DoubleU-NetPlus w/o TAG & 95.88 & 91.71 & 93.13 & 82.71\\
    & DoubleU-NetPlus w/o (TAM \& MKRC) & 96.46 & 91.81 & 94.01 & 84.45\\
    & DoubleU-NetPlus w/o (TAM \& MKRC \& TAG) & 96.93 &  92.17 & 94.20 & 83.58\\
    & DoubleU-NetPlus & 96.90 &  \textbf{92.47} & \textbf{94.30} & \textbf{84.71}\\
    
   \midrule
    CVCclinicDB & Baseline & 97.36 & 91.85 & 95.07  & 90.87\\
    & DoubleU-NetPlus w/o MKRC & 98.92 & 95.72 & 97.83 & 94.51\\
    & DoubleU-NetPlus w/o TAM & 98.89 & \textbf{96.12} & \textbf{98.05} & \textbf{95.52}\\
    & DoubleU-NetPlus w/o TAG & \textbf{98.96} &  95.72 & 97.83 & 94.51\\
    & DoubleU-NetPlus w/o (TAM \& MKRC) & 98.89 &  95.68 & 97.78 & 93.27\\
    & DoubleU-NetPlus w/o (TAM \& MKRC \& TAG) & 99.02 & 95.83 & 97.97 & 96.08\\
    & DoubleU-NetPlus & 97.96 &  93.87 & 96.40 & 95.12\\
   
   \midrule
    2018 DSB & Baseline & 96.84 &  78.55 & 92.09 & 85.60 \\
    & DoubleU-NetPlus w/o MKRC & 96.39 & 83.10 & 91.32 & 83.91\\
    & DoubleU-NetPlus w/o TAM & \textbf{99.03} & 82.67 & 90.84 & \textbf{90.57}\\
    & DoubleU-NetPlus w/o TAG & 97.58 & 83.29 & 89.21 & 85.73\\
    & DoubleU-NetPlus w/o (TAM \& MKRC) & 96.46 & 80.63 & 88.57 & 82.18\\
    & DoubleU-NetPlus w/o (TAM \& MKRC \& TAG) & 94.62 & 90.29 & 92.48 & 86.88\\
    & DoubleU-NetPlus & 98.82 &  \textbf{95.64} & \textbf{95.76} & 90.29\\
    
   \midrule
    ISBI 2012 & Baseline & 99.53 & 81.58 & 94.30 & 89.38\\
    & DoubleU-NetPlus w/o MKRC & 99.73 & 82.35 & 96.99 & 94.17\\
    & DoubleU-NetPlus w/o TAM & 99.61 & 82.31 & 96.72 & 93.66\\
    & DoubleU-NetPlus w/o TAG & 99.69 &  82.12 & 96.79 & 93.79\\
    & DoubleU-NetPlus w/o (TAM \& MKRC) & 99.69 & 82.04 & 96.66 & 93.54\\
    & DoubleU-NetPlus w/o (TAM \& MKRC \& TAG) & 99.61 & 82.27 & 96.62 & 93.47\\
    & DoubleU-NetPlus &  \textbf{99.75} &  \textbf{88.62} & \textbf{97.10} &  \textbf{94.38}\\
   \botrule
  \end{tabular*}
  \label{tab:ablation}
\end{table*}

\begin{table}[!htb]
  \footnotesize
  \caption{$P$-values between proposed DoubleU-NetPlus and other SOTA methods on different evaluation metrics.}
  \begin{tabular}{c c c}
   \toprule
    Pair & DICE & mIoU\\
   \midrule
    U-Net vs. DoubleU-NetPlus & 0.0101 & 0.0091\\
    U-Net++ vs. DoubleU-NetPlus & 0.0074 & 0.0231\\
    Att.U-Net vs. DoubleU-NetPlus & 0.0455 & 0.0364\\
    MultiResU-Net vs. DoubleU-NetPlus & 0.0432 & 0.0451\\
    FANet vs. DoubleU-NetPlus & 0.0003 & 0.0078\\
    DoubleU-Net vs. DoubleU-NetPlus & 0.0010 & 0.0185\\
    Sharp U-Net vs. DoubleU-NetPlus & 0.0114 & 0.0326\\
   \bottomrule
  \end{tabular}
  \label{tab:parametric_paired_t_test}
\end{table}

\subsubsection{Statistical significance test}

To statistically investigate the performance of the proposed DoubleU-NetPlus over other SOTA segmentation methods on different quantitative metrics, we conduct paired sample t-tests between the DICE and mIoU obtained by DoubleU-NetPlus and the DICE and mIoU obtained by other methods. A paired sample t-test is often used for comparing two methods on the same evaluation metric in the MIS domain~\cite{Msrf-net,Kiu-net,xue2021global,wang2021automated}. We perform the test on DICE and mIoU metrics mainly because these two are the most significant evaluation metrics in semantic image segmentation. It should be noted that we don't include the precision and recall metrics in the test because every compared method doesn't report these two metrics. A comparison was done with those methods which utilized all six datasets in their study or reported in the literature. A $p$-value less than 0.05 is considered as statistically significant and the paired-wise $p$-values are reported in Table ~\ref{tab:parametric_paired_t_test}. From Table  ~\ref{tab:parametric_paired_t_test}, it is clear that in all seven paired methods, the $p$-values are smaller than 0.05 for the DICE and mIoU metrics, which demonstrates that our proposed method achieved significantly improved results compared to seven other SOTA models.

\subsubsection{Ablation studies}

We have performed an extensive ablation study in each of the employed datasets to empirically verify some of our incorporated modules in the proposed DoubleU-NetPlus network. A baseline U-Net was used to benchmark the performance of various datasets that were used in our experiments. We investigate the baseline performance of the U-Net by training it with the same number of augmented images that were used to train the proposed DoubleU-NetPlus model and sequentially assess the performance with subsequent removal of the MKRC, TAM, TAG modules individually (TAM, and MKRC), and (TAM and MKRC and TAG), combinedly from the proposed architecture. The results of module removal on the BUSI and DRIVE dataset are demonstrated in Table ~\ref{tab:ablation}. It can be observed that the EfficientNetB7-based encoder backbone, TAM, MKRC, and TAG modules contribute significantly to the improvements in Dice score, mIoU, precision, and recall metric values.

\section{Conclusion}
Semantic segmentation of medical images is a key element in medical image analysis. This paper presents a robust deep learning-based MIS network named DoubleU-NetPlus equipped with several architectural modifications, mainly the integration of pre-trained EfficientNetB7 as a feature encoder backbone, a newly proposed multi-kernel residual convolution module, multi-scale feature re-calibrating SE-ASPP module, and a hybrid triple attention module at the bottleneck of each network. We also integrated attention-driven residual convolutions throughout the encoder and decoder part of the network. To capture salient regions with higher precision, we have integrated a novel triple attention gate module that focuses on the relevant regions and suppresses other irrelevant regions in the skip connections features. A combination of all these modules together captures high-level semantic and discriminative feature maps while preserving effective spatial information. Experimental results evaluated on six benchmark datasets of different modalities demonstrate the proposed model's superiority over SOTA segmentation methods in MIS tasks. We believe that DoubleU-NetPlus is a generic segmentation model and can be applied to similar 2D MIS tasks. One of the challenges in this architecture is its high number of trainable parameters. In the future, we plan to reduce the number of parameters and computational complexity. We also plan to adjust the design of the network to make it adaptable in the 3D image domain.

\subsection*{Author Contribution}

\textbf{Md. Rayhan Ahmed}: Conceptualization of this study,
Methodology, Formal analysis, Software, Writing - Original draft preparation, review, editing, and investigation.
\textbf{Adnan Ferdous Ashrafi}: Software, Visualization, Writing
- Original draft preparation, review, and editing. \textbf{Raihan
Uddin Ahmed}: Software, Validation, Writing - Original
draft preparation. \textbf{Swakkhar Shatabda}: Review, Validation, Editing. \textbf{A.K.M. Muzahidul Islam}: Review, Investigation. \textbf{Salekul Islam}: Review, Investigation.

\subsection*{Funding Information}
This research work did not receive any specific grant
from funding agencies in the public, commercial, or not-for-profit sectors.

\subsection*{Conflict of Interest}
The authors declare that they have no known competing financial interests or personal relationships that could have appeared to influence the work reported in this paper.

\subsection*{Data Availability}
The research has used publicly available datasets.

\bibliographystyle{sn-basic.bst}
% \printbibliography
\bibliography{sn-bibliography}% common bib file


\end{document}